\documentclass[12pt]{article}
\newcommand{\text}{\rm}
\begin{document}
\title{Large Dual Transformations and the Petrov-Diakonov
Representation of the Wilson Loop}
\author{L.~E.~Oxman\\ \\Instituto de F\'{\i}sica, Universidade Federal
Fluminense,\\
Campus da Praia Vermelha, Niter\'oi, 24210-340, RJ, Brazil.}
\date{\today}
\maketitle
\begin{abstract}

In this work, based on the Petrov-Diakonov representation of the Wilson loop
average $\bar{W}$ in the $SU(2)$ Yang-Mills theory, together with the Cho-Fadeev-Niemi
decomposition, we present a natural framework to discuss possible ideas
underlying confinement and ensembles of defects in the continuum.

In this language we show how for different ensembles the surface appearing in the 
Wess-Zumino term in $\bar{W}$ can be either decoupled or turned into a variable, 
to be summed together with gauge fields, defects and dual fields.

This is discussed in terms of the regularity properties imposed by the 
ensembles on the dual fields, thus precluding or enabling the possibility of
performing the large dual transformations that would be necessary to decouple
the initial surface.

\end{abstract}

\subsection*{Keywords} 

Duality in nonabelian gauge theories, monopoles and center vortices, confinement
vs. deconfinement.

\section{Introduction}

Nowadays, one of the most important and interesting open problems in Physics
corresponds to understanding quark confinement. Although quantum chromodynamics
is completely successful in describing high energy phenomena, where because of
asymptotic freedom the main characters are quarks and gluons, a theoretical
explanation for the confinement of these objects in colorless asymptotic states
is still lacking.

With regard to gluon confinement, an important line of research corresponds to
studying the effect of the Gribov horizon \cite{gqcd}  on the gluon propagator.
These ideas indicate that the inclusion of a Gribov-Zwanziger term in the pure
Yang-Mills action, as to avoid Gribov copies, leads to infrared suppressed gluon
and ghost propagators  \cite{Cucchieri:2007md}-\cite{refinement}. While the
absence of the pole in the gluon propagator would explain why gluons cannot
occupy asymptotic states, it is difficult to imagine an explanation for quark
confinement in this framework, as the infrared suppression could not produce
long range forces.

Therefore, among the possible frameworks for the confinement of (heavy) quarks
in pure Yang-Mills theory, those based on the inclusion of a nonperturbative
sector represented by magnetic defects become favored, and the problem turns out
to be the identification of defects, their associated phases, and how they can
imply an area law for the Wilson loop. Although these points have been studied
for many years now, a closed theoretical understanding is still lacking
\cite{greensite}-\cite{CG1}.

For example, in the mechanism of dual superconductivity \cite{3}-\cite{M},
\cite{polya}, the QCD vacuum is expected to behave as a superconductor of
chromomagnetic charges, which implies the confinement of chromoelectric charges,
in an analogous (dual) manner to what would happen with a type II
superconductor, where magnetic monopoles would be confined because of the
magnetic flux tube generated between them.

When implementing the Abelian projection \cite{ap}, monopoles can appear as
defects when a gauge fixing that diagonalizes a field that transforms in the
adjoint representation of $SU(N)$ is considered. 

Another possible manner to identify them is as defects when trying to implement
the Cho-Faddeev-Niemi (CFN) decomposition, with the advantage that in this case
no particular gauge fixing condition is invoked. For instance, the monopoles for
$SU(2)$ are defects of the local direction $\hat{n}$ used to decompose the
connection in color space (see \cite{cho-a}, \cite{cho2}-\cite{Shaba} and
references therein),
\begin{equation}
\vec{A}_\mu=A^{(n)}_\mu \hat{n}-\frac{1}{g} \hat{n}\times \partial_\mu \hat{n} +
\vec{X}^{(n)}_\mu
\makebox[.5in]{,}
\hat{n}\cdot\vec{X}^{(n)}_\mu=0 .
\label{dec}
\end{equation}

Besides monopoles, $Z(N)$ center vortices are also of great interest, as they
could explain the string tension dependence on the representation of the
subgroup $Z(N)$ of $SU(N)$ observed in the lattice ($N$-ality), a property that
cannot be explained by the isolated effect of monopoles. In addition, when
closed center vortices are included, an area law (confining phase) or perimeter
law (deconfining phase) has also been observed, depending on whether these
objects percolate or not \cite{debbio3}-\cite{quandt}. 

Moreover, strong correlations between monopoles and center vortices are
supported by recent results on the lattice, 
and they are quite promising in accommodating the different properties of the
confining phase \cite{AGG}-\cite{GKPSZ} (for a review, see also ref.
\cite{greensite}).

The aim of this work is to present a natural framework to discuss possible ideas
underlying confinement and ensembles of defects in the continuum.

In this regard, we have recently unified the description of monopoles and center
vortices \cite{lucho} as different types of defects of the complete local color
frame $\hat{n}_a$, $a = 1, 2, 3$ used in the Cho-Faddeev-Niemi decomposition of
the $SU(2)$ gauge fields; where $\hat{n}_3=\hat{n}$ and,
\begin{equation}
\vec{X}^{(n)}_\mu =X_\mu^1 \hat{n}_1+X_\mu^2 \hat{n}_2 .
\end{equation} 

When the element $\hat{n}$ contains monopole-like defects, localized on closed
strings, the elements $\hat{n}_1$, $\hat{n}_2$ inevitably contain defects on
open surfaces, and these can correspond to Dirac worldsheets or to pairs of
center vortex worldsheets, attached to the monopoles. When we go close to and
around an open center vortex (Dirac) worldsheet, $\hat{n}_1$ and $\hat{n}_2$
rotate once (twice), corresponding to the flux $2\pi/g$ ($4\pi/g$) carried by
them.

In this manner, additional singular terms in the Yang-Mills action appear, due
to the fact that derivatives do not commute when defects are present. These are
either localized on Dirac worldsheets or on thin center vortices.

In fact, these singular terms were missing in previous literature about the
Cho-Faddeev-Niemi decomposition. In this respect, we would like to point out
that effective Skyrme models have been constructed in terms of $\hat{n}$
\cite{FN,Shaba}, \cite{cho1,F,kondo.sky}, guided by the decomposition in eq.
(\ref{dec}). Then, although they capture information about monopoles without
reference to unobservable Dirac worldsheets, as expected in a well-defined
effective model, the information about center vortices in the $\hat{n}_1$,
$\hat{n}_2$ sector is lost in this heuristic process (for a discussion, see
refs. \cite{lucho,AML}).

In this article we will first give a representation for the Wilson loop average
$\bar{W}$ in the $SU(2)$ Yang-Mills theory, similar to the one in refs.
\cite{cho2}, \cite{kondo3}, but including the singular terms for the monopole
and the center vortex sectors. For this purpose we will use the Petrov-Diakonov
(PD) representation of the Wilson loop \cite{P-D}-\cite{diako}, as the natural
variables here are those used in the Cho-Faddeev-Niemi decomposition
\cite{cho2}.

In particular, for a given gauge field $\vec{A}_\mu$, the Wilson loop order
parameter $W({\cal C})$ can be written as an integral over $U\in SU(2)$
containing an Abelian looking integrand that depends on $A^{(n)}_\mu$, the field
that appears in the decomposition of $\vec{A}_\mu$ with respect to the local
frame induced by $U$ (for a brief review, see section \S \ref{PDrep}). The
important point is that this representation also includes a Wess-Zumino term,
concentrated on a ``Wilson surface'' $S({{\cal C}})$, whose border is the Wilson
loop ${\cal C}$, although the usual representation for $W({\cal C})$ contains no
reference to a surface. 

In the Petrov-Diakonov representation any surface $S({{\cal C}})$ can be used,
up to singular situations where it passes over the monopoles
\cite{diako-petro-reply}. This raises the problem of how to deal with this
arbitrary surface in the average over fields and ensembles of defects. In ref.
\cite{polyakov1}, this kind of problem has been discussed in the context of
compact $QED(3)$ and $QED(4)$.

Using our representation for $\bar{W}$, we will discuss here how monopole and
center vortex ensembles can render the surface appearing in the Wess-Zumino term
a variable, to be summed together with gauge fields, 
defects and dual fields. 
This occurs when the regularity properties imposed by the associated
physical phases on the dual fields preclude the implementation of large dual
field transformations in the path integral, a necessary step that should be
considered in order to decouple the initial Wilson surface and show it is an unobservable
object. 

In general, using our arguments in $3D$ or $4D$, prior to the ensemble
integration, we will obtain a representation evidencing the decoupling of the initial Wilson
surface or its replacement by a  ``Wilson surface variable'', depending on the assumed 
closure properties for the dual fields.

In $3D$, as center vortices are string-like, we will also be able to propose the
general form of an effective action describing the interaction between gauge,
vortex and dual fields, as well as Wilson surfaces. Therefore, the relationship
between deconfining/confining ensembles and closure/nonclosure properties of
the large dual transformations will be clear in this case. 

Of course, which is the correct ensemble of defects associated with Yang-Mills
theories is the fundamental part of the problem of confinement. In particular,
how can the dressing of thin defects lead to dimensional parameters
characterizing thick objects that condense. This is outside the scope of this
article, which is organized in the following manner. 

In sections \S \ref{m-cho} and \S \ref{PDrep}, we review how to describe
monopoles and center vortices in terms of the defects of the complete local
color frame used to decompose the gauge fields,
as well as the Petrov-Diakonov representation of the Wilson loop $W$. Section \S
\ref{Wl-av} is dedicated to a brief discussion of the representation for the
average $\bar{W}$, including a general ensemble of monopoles and center
vortices. 

In \S \ref{unc-c}, we discuss the arbitrary Wilson surface $S({\cal C})$ in
connection with the integrand of $\bar{W}$.
In section \S \ref{sym-clos}, we present possible effective models that describe
chains of correlated monopoles and center vortices, and discuss how they could
preclude the implementation of large dual changes of variables.

In \S \ref{dec-obs}, we show how to decouple the Wilson surface $S({\cal C})$ in
favor of its border, in the case where the dual fields are closed under large
dual transformations. In the opposite case, we show how $S({\cal C})$ is replaced by a 
Wilson surface variable, also including a discussion of generalized multivalued dual
fields in continuum $4D$ theories. 

Finally, we present our conclusions in section \S \ref{conc}.

\section{Defects of the local color frame}
\label{m-cho}

When studying Abelian projection scenarios, the gauge fields are generally
separated into ``diagonal'' fields, living in the Cartan subalgebra of $SU(N)$,
and ``off-diagonal'' charged fields. 
For instance, in the case of $SU(2)$, the uncharged sector can be chosen along
the $\hat{e}_3$ direction in color space, while the components along $\hat{e}_1$
and $\hat{e}_2$ correspond to the charged sector.

In the CFN decomposition, this separation into charged and uncharged sectors is
also implemented, with the advantage that it is naturally done along a general
$\hat{n}_3 = \hat{n}$ local direction in color space.

In ref. \cite{lucho}, we have unified monopoles and center vortex worldsheets as
different classes of defects in the local color frame $\hat{n}_a=R\, \hat{e}_a$,
$R\in$ SO(3), used in the CFN decomposition. While it is well-known that
monopole-like defects are associated with a nontrivial $\Pi_2$ for the space of
directions $\hat{n}$, we can also think of thin center vortices as the natural
defects of a frame, due to the nontrivial fundamental group $\Pi_1=Z(2)$ of
$SO(3)$.

The possibility of matching general nontrivial configurations containing
mo\-no\-poles and center vortices is evidenced by parametrizing the gauge fields
in terms of the CFN decomposition, based on a class of frames $\hat{n}_a$,
\begin{equation}
(VU) T^a (VU)^{-1}= \hat{n}_a\cdot\vec{T}
\makebox[.5in]{,}
\hat{n}_a=R(VU)\hat{e}_a
\label{nVU}
\end{equation}
where $U$ is single-valued along any closed loop, defining a frame $\hat{m}_a$,
\begin{equation}
U T^a U^{-1}= \hat{m}_a\cdot\vec{T}
\makebox[.5in]{,}
\hat{m}_a=R(U)\hat{e}_a,
\end{equation}
such that $\hat{m}_3=\hat{m}$ is a topologically nontrivial mapping that encodes
the mon\-o\-pole sector.
The $V$ part is multivalued and enables the description of the center vortex
sector. 

Let us consider, for example, a gauge field whose decomposition is given by,
\begin{equation}
\vec{a}_\mu \cdot\vec{T}=-(C^{(n)}_\mu \hat{n}+\frac{1}{g}\hat{n} \times
\partial_\mu \hat{n})\cdot\vec{T}
\label{Ugauge}
\makebox[.5in]{,}
C^{(n)}_\mu=-\frac{1}{g}\hat{n}_1\cdot\partial_\mu \hat{n}_2 .
\end{equation}
In the case where $V\equiv I$, and taking $U=\bar{U}= e^{-i\varphi T_3}
e^{-i\theta T_2}  e^{+i\varphi T_3}$, where $\varphi$ and $\theta$ are the polar
angles defining $\hat{r}$, eq. (\ref{Ugauge}) corresponds to a nontrivial
``gauge'' transformation $\frac{i}{g} \bar{U}\partial_\mu \bar{U}^{-1}$
introducing an anti-monopole \cite{cho2}. Note that no singularity is present at
$\theta \approx 0$, where $\bar{U}\approx I$. The Dirac string is placed at
$\theta =\pi$; when we go close and around the negative $z$-axis, the elements
$\hat{n}_1$, $\hat{n}_2$ rotate twice. A monopole is obtained with the
replacement $\theta \rightarrow \pi-\theta$, $\varphi \rightarrow\varphi +\pi$.

More generally, a field decomposed according to eq. (\ref{dec}), with $V\equiv
I$, can be written as a nontrivial transformation of a regular background
$\vec{\cal A}_\mu$, 
\begin{equation}
\vec{A}_\mu\cdot\vec{T}=\vec{\cal A}_\mu^{\bar{U}}\cdot\vec{T}=\bar{U} \vec{\cal
A}_\mu\cdot\vec{T} \bar{U}^{-1}+\frac{i}{g} \bar{U}\partial_\mu \bar{U}^{-1}.
\label{m-gauge}
\end{equation}
As is well-known, the field strength for $\vec{\cal A}_\mu^{\bar{U}}$ is,
\begin{eqnarray}
\vec{\cal F}_{\mu \nu}^{\bar{U}}\cdot\vec{T} &=&\bar{U} \vec{\cal F}_{\mu
\nu}\cdot\vec{T} \bar{U}^{-1}+\frac{i}{g}\bar{U} [\partial_\mu,\partial_\nu]
\bar{U}^{-1}.
\label{m-c}
\end{eqnarray}
That is, the fields $\vec{A}_\mu$ and $\vec{\cal A}_\mu^{\bar{U}}$ are not
physically equivalent, because of the second term in eq. (\ref{m-c}) which is
concentrated on a Dirac worldsheet, namely, the two-dimensional surface where $\bar{U}$ is singular.

Now, by considering in eq. (\ref{Ugauge}) a local frame defined by $U\equiv I$
and $V=\bar{V}=e^{i\varphi \, T_3}$, we obtain,
\begin{equation}
\vec{a}_\mu \cdot\vec{T} =\frac{1}{g} \partial_\mu \varphi\, \delta^{a3} T^a,
\end{equation}
that is, a thin center vortex placed on the two-dimensional surface formed by
the $z$-axis, for every Euclidean time.
As the transformation $\bar{V}=e^{i\varphi \, T_3}$ is not single-valued, we
have, 
\begin{equation}
\frac{1}{g} \partial_\mu \varphi\, \delta^{a3} T^a 
=\frac{i}{g}\bar{V}\partial_\mu \bar{V}^{-1} -~{\rm ideal~vortex},
\end{equation}
where the additional term (the so called ideal vortex) is localized on the
three-volume where the transformation is discontinuous. For a general discussion
of thin and ideal center vortices in the continuum, see refs.
\cite{engelhardt1,reinhardt}.  Then, unlike monopoles, center vortices can only
be written in the form 
$\frac{i}{g}\bar{V}\partial_\mu \bar{V}^{-1}$ on a region outside the above
mentioned three-volume.  

Furthermore, if on the monopole ansatz after eq. (\ref{Ugauge}), $V\equiv I$
were replaced by $\bar{V}= e^{-i\varphi\, \hat{m}\cdot\vec{T}}$, we would have
$\bar{V}\bar{U}=e^{-i\varphi T_3} e^{-i\theta T_2}$.
Then, instead of a monopole attached to a Dirac worldsheet placed at
$\theta=\pi$, one attached to a pair of center vortices at 
$\theta=0$ and $\theta=\pi$ would be obtained. In this case, when we go close
and around the positive and negative $z$-axis,
the elements $\hat{n}_1$, $\hat{n}_2$ rotate once, with different orientations.
In general, any configuration containing 
monopoles and center vortices (correlated or not) can be written in terms of
three Euler angles $\bar{V}\bar{U}= e^{-i\alpha T_3} e^{-i\beta T_2} 
e^{+i(\alpha -\gamma)T_3}$, that corresponds to a single-valued 
$\bar{U}= e^{-i\alpha T_3} e^{-i\beta T_2}  e^{+i\alpha T_3}$, and a rotation
$\bar{V}= e^{-i\gamma\, \hat{m}\cdot\vec{T}}=\bar{U} e^{-i\gamma\,
T_3}\bar{U}^{-1}$,
leaving $\hat{m}=\hat{n}$ fixed.

\section{Petrov-Diakonov representation}
\label{PDrep}

The usual representation for the nonabelian Wilson loop order parameter is given
by,
\begin{equation}
W({\cal C})= (1/2) tr\, P \exp (ig\oint dx_\mu \vec{A}_\mu \cdot\vec{T}).
\end{equation}

There is an alternative representation, due to Petrov and Diakonov
\cite{P-D}-\cite{diako}. For quarks in the fundamental representation, it is
given by,
\begin{equation}
W({\cal C})=(1/2)\int [{\cal D}U(\tau)]\, e^{\frac{i}{2}g\int_{0}^{1} d\tau\,
tr\,[ \tau^3(U^{-1} A U+\frac{i}{g} U^{-1} \frac{d~}{d\tau} U)]},
\label{W-PD}
\end{equation}
\begin{equation}
A(\tau)=\frac{dx_\mu}{d\tau} \vec{A}_\mu\cdot\vec{T}.
\label{Atau}
\end{equation}
Here, the Wilson loop ${\cal C}$ has been parametrized as, $x_\mu=x_\mu(\tau)$,
$\tau\in[0,1]$, $x_\mu(0)=x_\mu(1)$. 
The integration measure is,
\begin{equation}
\int [{\cal D}U(\tau)]\,= \int dU \int_{U(0)=U}^{U(1)=U} {\cal D}U(\tau)\, ,
\label{measure-PD}
\end{equation}
which means that the functional integral is done over $U$-transformations that
are single-valued along the Wilson loop.

Considering that on a given loop it is always possible to write,
\begin{equation}
A(u)=\frac{i}{g} Q^{-1} \frac{d~}{d\tau}Q
\makebox[.5in]{,}
Q(u)=\exp \left( -ig \int_{0}^u du' A(u')\right),
\end{equation}
it results \cite{P-D}-\cite{diako},
\begin{eqnarray}
W({\cal C})&=&(1/2)\int dU \int_{U(0)=U}^{U(1)=U} {\cal D}U(\tau)\,
e^{\frac{i}{2} \int_{0}^{1} d\tau\,
tr\,[ \tau^3(i (QU)^{-1} \frac{d~}{d\tau}(QU))]},\nonumber \\
&=&(1/2)\sum_{\alpha}D^{(1/2)}_{\alpha \alpha}(Q^{-1}(1)Q(0)).
\label{traceD} 
\end{eqnarray}
Of course the Wilson variable generally takes a nontrivial value, that is,
$Q(1)$ is generally not $Q(0)=1$.

To see how these expressions work, let us recall that closed center vortices are
usually defined as defects in the connection such that $W({\cal C})$ changes
sign when the defect is linked, and is otherwise left unchanged.

As is well-known, considering a line $x(\tau)$ which lives on a simply connected
region outside a closed vortex, where it is possible to write
$\vec{A}_\mu=\vec{\cal A}^{\bar{V}\bar{U}}_\mu$, and then taking the limit where
their endpoints are joined to form the loop ${\cal C}$, the usual representation
for $W({\cal C})$ gives $e^{iq\pi}W_{{\cal A}}({\cal C})$, where $W_{{\cal
A}}({\cal C})$ is the Wilson loop for the field ${\cal A}_\mu$.

Now, we can use the PD representation. From eq. (\ref{Atau}), we have,
\begin{equation}
A(\tau)=\frac{dx_\mu}{d\tau} \vec{A}_\mu\cdot\vec{T}=[(\bar{V}\bar{U}){\cal
A}(\tau)(\bar{V}\bar{U})^{-1}+\frac{i}{g}(\bar{V}\bar{U})\frac{d}{d\tau}(\bar{V}
\bar{U})^{-1}],
\end{equation}
where we have defined ${\cal A}(\tau)=\frac{dx_\mu}{d\tau} \vec{{\cal
A}}_\mu\cdot\vec{T}$.
Recalling that on the loop we can always write ${\cal A}(\tau)=\frac{i}{g} {\cal
Q}^{-1} \frac{d~}{d\tau}{\cal Q}$,
we get $Q={\cal Q}\, \bar{U}^{-1}\bar{V}^{-1}$. Then, using in eq.
(\ref{traceD}) the cyclic property of the trace,
and considering that $D^{(1/2)}$ is an odd function, the previous result is
reobtained,
\begin{equation}
W({\cal C})=(1/2)\sum_{\alpha}D^{(1/2)}_{\alpha \alpha}({\cal
Q}(0)\bar{U}^{-1}_i\bar{V}^{-1}_i\bar{V}_f
\bar{U}_f {\cal Q}^{-1}(1))=e^{iq\pi}W_{{\cal A}}({\cal C}).
\end{equation}

It is important to underline that the second part in the exponent of eq.
(\ref{W-PD}) is a Wess-Zumino term, and can be rewritten not in terms of a line
but in terms of a surface integral \cite{P-D}-\cite{diako}. Therefore, in
general, we have,
\begin{eqnarray}
W({\cal C})&=&(1/2)\int [{\cal D}U(\tau,\xi)]\,
e^{i\frac{g}{2} \int d^4x\, s_{\mu \nu} (f^{(m)}_{\mu \nu}+h^{(m)}_{\mu \nu})},
\label{Wtausigma}
\end{eqnarray}
where the source $s_{\mu \nu}$ is concentrated on a surface $S(\cal C)$ whose
border is the Wilson loop 
${\cal C}$, and is constructed by requiring $\int d^4x\, s_{\mu \nu}
(f^{(m)}_{\mu \nu}+h^{(m)}_{\mu \nu})$ to be the flux of $f^{(m)}_{\mu
\nu}+h^{(m)}_{\mu \nu}$ through $S(\cal C)$. This surface can be parametrized by
$x(\tau, \xi)$, and $s_{\mu \nu}$ must satisfy,
\begin{equation}
j_\mu({\cal C})=\epsilon_{\mu \nu \rho \sigma} \partial_\nu s_{\rho \sigma},
\makebox[.5in]{,}
j_\mu({\cal C})=\int d\tau \, \frac{dx_\mu}{d\tau} \delta (x-x(\tau)),
\label{SC}
\end{equation}
where $x(\tau)=x(\tau, 1)$ is a parametrization of ${\cal C}$. In eq.
(\ref{Wtausigma}), we also have, 
\begin{equation}
f^{(m)}_{\mu \nu}=f^{(m)}_{\mu \nu}=\epsilon_{\mu \nu \rho \sigma}\partial_\nu
A^{(m)}_\sigma  \makebox[.5in]{,}
h^{(m)}_{\mu \nu}=-\frac{1}{2g} \epsilon_{\mu \nu \rho \sigma}
\hat{m}\cdot(\partial_\rho \hat{m} \times \partial_\sigma \hat{m}),
\end{equation}
where the connection is decomposed by using a frame $\hat{m}_a$, defined on
$S(\cal C)$, and induced by $U(\tau,\xi)$, namely,
\begin{equation}
U T_a U^{-1}=\hat{m}_a\cdot\vec{T},
\end{equation}
\begin{equation}
\vec{A}_\mu=A^{(m)}_\mu \hat{m}-\frac{1}{g} \hat{m}\times \partial_\mu \hat{m} +
\vec{X}^{(m)}_\mu.
\end{equation}
We also note that the possibility of writing,
\begin{equation}
\int_{0}^{1} d\tau\, \frac{i}{g} tr\,[ \tau^3 U^{-1} \frac{d~}{du}U]=\int d^4x\,
s_{\mu \nu}\, h^{(m)}_{\mu \nu},
\end{equation}
depends on the single valuedness of $U(\tau)$ (see ref.
\cite{diako-petro-reply}). This condition is met precisely because of the
integration measure in eq. (\ref{measure-PD}). 

\section{Wilson loop average}
\label{Wl-av}

Now we will work with thin objects defined on the whole Euclidean spacetime,
taking into account the singular terms arising from the color frame defects. Let
us consider the Wilson loop average,
\begin{eqnarray}
\bar{W}({\cal C}) &=& \frac{1}{2{\cal N}} \int [{\cal D}\vec{A}] F_{gf}\,
e^{-S_{YM}[\vec{A}]}\, tr\, P \exp (ig\oint dx_\mu \vec{A}_\mu \cdot\vec{T}),
\end{eqnarray}
\begin{equation}
{\cal N}=\int [{\cal D}\vec{A}] F_{gf}\, e^{-S_{YM}[\vec{A}]},
\end{equation}
where $F_{gf}$ is the part of the measure that fixes the gauge, including in
general  auxiliary fields.

Using the PD representation, we have,
\begin{equation}
\bar{W}({\cal C})=\frac{1}{2{\cal N}} \int [{\cal D}\vec{A}][{\cal
D}U(\tau,\xi)] F_{gf}\, e^{-S_{YM}[\vec{A}]}\,
e^{\frac{i}{2}g \int d^4 x\, s_{\mu \nu} (f^{(m)}_{\mu \nu}+h^{(m)}_{\mu \nu})},
\end{equation}
In fact, as we are interested in discussing the Wilson loop globally, for any
closed loop and any associated surface, we will have to consider the extension
$U(x)$, defined on the whole Euclidean spacetime, up to possible singularities,
such that $U(x(\tau,\xi))=U(\tau,\xi)$.

Now, as the Wilson loop is written in terms of the CFN variables, it is
convenient to change to these variables in the path-integral \cite{cho2,kondo3}.
The procedure is to include the integration over the extended $U$'s, which
amounts to introducing a product of group volumes, and then performing a change
(with unit Jacobian) to the variables $A^{(m)}_\mu$, $\vec{X}^{(m)}_\mu$ ($m=1,2$) in
the decomposition of $\vec{A}_\mu$ with respect to the basis induced by $U(x)$.

An important point to be underlined is that after the change,
$\vec{A}_\mu$-con\-fig\-u\-ra\-tions containing monopoles  
will be represented by $U$'s inducing frames with monopole-like defects in
$\hat{m}$. In addition, as $U$-configurations
are single-valued, thin center vortices will be manifested as defects in the
components of the charged fields $\vec{X}^{(m)}_\mu$. For convenience, the
ensemble integration over these defects can be replaced by the integration over 
a $V$-sector, that according to eq. (\ref{nVU}) rotates $\hat{m}_1$, $\hat{m}_2$
to $\hat{n}_1$, $\hat{n}_2$, leaving $\hat{m}=\hat{n}$ fixed. This is done in
order to identify monopoles and center vortices with singular frames. Then, we
have,
\begin{equation}
\bar{W}({\cal C}) 
= \frac{1}{2{\cal M}}\int [{\cal D}A][{\cal D}X][{\cal D}U][{\cal D}V] F_{gf}\, 
e^{-S_{YM}[\hat{n}_a, A^{(n)},X^{(n)}]}\,
e^{\frac{i}{2} g \int d^4 x\, s_{\mu \nu} (f^{(n)}_{\mu \nu}+h^{(n)}_{\mu
\nu})}, \nonumber
\end{equation}
\begin{equation}
{\cal M}=\int [{\cal D}A][{\cal D}X][{\cal D}U][{\cal D}V] F_{gf}\,
e^{-S_{YM}[\hat{n}_a,A^{(n)},X^{(n)}]}.
\end{equation}

A fundamental ingredient to be taken into account is regarding the nontrivial
singular terms associated with the frame defects.
In ref. \cite{lucho}, we have identified two types, which were missing in the
field strength tensor computed in refs. \cite{cho2}-\cite{cho5}. The first one
depends on defects of the third component $\hat{n}_3\equiv \hat{n}$, and occurs
in the charged sector of the field strength tensor. In ref. \cite{lucho}, this
type of term has been nullified by considering $\hat{n}$-configurations that
have at most monopole defects. In this case, $S_{YM}$ results,
\begin{eqnarray}
S_{YM}&=&\int d^4x\, \left[ \frac{1}{4} (f^{(n)}_{\mu \nu} +h^{(n)}_{\mu
\nu}+k_{\mu \nu})^2 + \frac{1}{2} \bar{g}^{\mu \nu} g^{\mu \nu}\right],
\label{SM}
\end{eqnarray}
where,
\begin{equation}
g^{\mu \nu}=\epsilon^{\mu \nu \rho
\sigma}[\partial_\rho+ig(A^{(n)}_\rho+C^{(n)}_\rho)]\Phi_\sigma,
\makebox[.5in]{,}
C^{(n)}_\mu =-\frac{1}{g} \hat{n}_1.\partial_\mu \hat{n}_2.
\end{equation}
\begin{equation}
\Phi_\mu=\frac{1}{\sqrt{2}}(X^1_\mu+iX^2_\mu)
\makebox[.5in]{,}
k_{\mu \nu}=\frac{g}{2i}\epsilon_{\mu \nu \rho \sigma}(\bar{\Phi}_\rho
\Phi_\sigma-\Phi_\rho \bar{\Phi}_\sigma),
\end{equation}
\begin{equation}
f^{(n)}_{\mu \nu}=\epsilon_{\mu \nu \rho \sigma}\partial_\rho A^{(n)}_\sigma
\makebox[.5in]{,}
h^{(n)}_{\mu \nu}=-\frac{1}{2g} \epsilon_{\mu \nu \rho \sigma} \,
\hat{n}\cdot(\partial_\rho \hat{n} \times \partial_\sigma \hat{n}).
\end{equation}
The second type occurs when trying to express the monopole part $h^{(n)}_{\mu
\nu}$ of the dual field strength 
in terms of the monopole potential $C^{(n)}_\mu$. In this case, we obtain,
\begin{equation}
h_{\mu \nu}=\tilde{h}^{(n)}_{\mu \nu}+ d^{(n)}_{\mu \nu},
\makebox[.5in]{,}
\label{mv-sep}
\tilde{h}^{(n)}_{\mu \nu}=\epsilon_{\mu \nu \rho \sigma}\partial_\rho
C^{(n)}_\sigma ,
\end{equation}
where the singular terms $d^{(n)}_{\mu \nu}$ are concentrated on the frame
defects. If not for this difference, the
surface integral in the Wess-Zumino term of the PD representation could be
converted into a line integral.

Now we can proceed as we did for the partition function in ref. \cite{lucho}.
Introducing real and complex lagrange multipliers, $\lambda_{\mu \nu}$ and
$\Lambda_{\mu \nu}$, we get,
\begin{eqnarray}
\bar{W}({\cal C}) &=& \frac{1}{2{\cal M}} \int [{\cal D}\lambda][{\cal
D}\Psi][{\cal D}U][{\cal D}V]\, e^{-S_c-\int d^4x\, \frac{1}{4}\lambda_{\mu \nu}
\lambda_{\mu \nu}}\times\nonumber \\
&&\times e^{i\int d^4x\, [\frac{1}{2}\lambda_{\mu \nu}(f^{(n)}_{\mu
\nu}+h^{(n)}_{\mu \nu}+k_{\mu \nu})- 
J^\mu_c(A^{(n)}_\mu+C^{(n)}_\mu) +\frac{g}{2}  s_{\mu \nu}
(f^{(n)}_{\mu\nu}+h^{(n)}_{\mu \nu})]}.
\end{eqnarray}
where we have defined, $[{\cal D}\Psi]=[{\cal D}A^{(n)}][{\cal D}\Phi][{\cal
D}\Lambda] \tilde{F}_{gf}$. Here, we have the action for the charged fields,
\begin{equation}
S_c=\int d^4x\, \left[\frac{1}{2}\bar{\Lambda}^{\mu \nu} \Lambda^{\mu
\nu}-\frac{i}{2} (\bar{\Lambda}^{\mu \nu} 
\epsilon^{\mu \nu \rho \sigma}\partial_\rho \Phi_\sigma + {\Lambda}^{\mu \nu} 
\epsilon^{\mu \nu \rho \sigma}\partial_\rho \bar{\Phi}_\sigma)\right],
\label{ch-action}
\end{equation}
minimally coupled to the $U(1)$ color current $J^\mu_c=J^\mu +K^\mu$, 
\begin{equation}
J^\mu = -\frac{i}{2}\, g \epsilon^{\mu \nu \rho \sigma} \bar{\Lambda}_{\nu
\rho}\Phi_\sigma + \frac{i}{2}\, g \epsilon^{\mu \nu \rho \sigma} {\Lambda}_{\nu
\rho}\bar{\Phi}_\sigma.
\label{Jlambda}
\end{equation}
The terms $K_\mu$ and $\tilde{F}_{gf}$ appear when fixing an extended maximally
Abelian gauge,
\begin{equation}
\partial_\mu (A^{(n)}_\mu+C^{(n)}_\mu)=0.
\label{lorentz}
\end{equation}
\begin{equation}
[\partial_\mu +ig(A^{(n)}_\mu+C^{(n)}_\mu)] \Phi_\mu =0
\makebox[.5in]{,}
[\partial_\mu -ig(A^{(n)}_\mu+C^{(n)}_\mu)] \bar{\Phi}_\mu =0.
\label{MAG}
\end{equation}
More precisely, 
\begin{equation}
F_{gf}=\tilde{F}_{gf}\,  e^{-i\int d^4x\, (A^{(n)}_\mu+C^{(n)}_\mu)K^\mu} ,
\end{equation}
where $\tilde{F}_{gf}$ is independent of $A^{(n)}_\mu$, and contains the
integration measure
for lagrange multipliers, ghosts and auxiliary fields, while $K^\mu$ depends on
these fields, as well as on $\Phi_\mu$.

Because of the $A^{(n)}_\mu$ path integration, a constraint is implicit here, 
\begin{equation}
J_c^\mu=\frac{1}{2}\epsilon_{\mu \nu \rho \sigma}\partial_\nu (\lambda_{\rho
\sigma}+g s_{\rho \sigma}),
\label{j-map}
\end{equation}
so that we finally get,
\begin{eqnarray}
\bar{W}({\cal C})
&= &\int [{\cal D}\lambda][{\cal D}\Psi][{\cal D}U][{\cal D}V]\, e^{-S_c-\int
d^4x\, \frac{1}{4}\lambda_{\mu \nu} \lambda_{\mu \nu}}\times\nonumber \\
&&\times e^{i\int d^4x\, \{ (\frac{1}{2}\epsilon_{\mu \nu \rho \sigma}
\partial_\nu (\lambda_{\rho \sigma}+g
s_{\rho \sigma})-
J^c_\mu ) A^{(n)}_\mu +\frac{1}{2}\lambda_{\mu \nu} k_{\mu
\nu}+\frac{1}{2}(\lambda_{\mu \nu}+g s_{\mu \nu})d^{(n)}_{\mu \nu}\} }.\nonumber
\\
\label{WYMb}
\end{eqnarray}

It will also be convenient to discuss the representation in $3D$, derived by
following the same steps, namely,
\begin{eqnarray}
\bar{W}({\cal C}) &=&  \int [{\cal D}\lambda][{\cal D}\Psi][{\cal D}U][{\cal
D}V]\, e^{-S_c-\int d^3x\, \frac{1}{2}\lambda_{\mu}
\lambda_{\mu}}\times\nonumber \\
&&\times e^{i\int d^3x\, \{ (\epsilon_{\mu \nu \rho} \partial_\nu
(\lambda_{\rho}+\frac{g}{2}\, s_{\rho})-
J^c_\mu ) A^{(n)}_\mu +\lambda_{\mu} k_{\mu}+(\lambda_{\mu }+\frac{g}{2}\,
s_{\mu})d^{(n)}_{\mu}\}},
\label{W3d}
\end{eqnarray}
\begin{equation}
S_c=\int d^3x\, \left[\bar{\Lambda}^{\mu} \Lambda^{\mu}-i (\bar{\Lambda}^{\mu } 
\epsilon^{\mu \nu \rho }\partial_\nu \Phi_\rho + {\Lambda}^{\mu } 
\epsilon^{\mu \nu \rho }\partial_\nu \bar{\Phi}_\rho)\right].
\end{equation}
In the total charge current $J_c^\mu = J^\mu +K^\mu$, the term $K^\mu$ receives
contributions from the charged fields of the gauge fixing sector and,
\begin{equation}
J^\mu= ig \epsilon^{\mu \nu \rho} \bar{\Lambda}_{\nu }\Phi_\rho - i
g\epsilon^{\mu \nu \rho } {\Lambda}_{\nu }\bar{\Phi}_\rho
\makebox[.5in]{,}
k_\mu=\frac{g}{2i}\epsilon_{\mu \nu \rho}(\bar{\Phi}_\nu \Phi_\rho- \Phi_\nu
\bar{\Phi}_\rho).
\end{equation}
The source $s_{\mu}$ is concentrated on $S({\cal C})$, and is such that $\int
d^3x\, s_\mu (f_\mu+h_\mu)$ gives the flux of $(f_\mu+h_\mu)$. Also in eq.
(\ref{W3d}), we have the implicit constraint,
\begin{equation}
J^c_\mu=\epsilon_{\mu \nu \rho} \partial_\nu (\lambda_{\rho}+\frac{g}{2}\,
s_{\rho})
\makebox[.5in]{,}
\epsilon_{\mu \nu \rho} \partial_\nu s_\rho =j_\mu ({\cal C}).
\label{const-3d}
\end{equation}
Finally, $d^{(n)}_{\mu}$ is concentrated on the defects and is obtained from,
\begin{equation}
h^{(n)}_{\mu}=\tilde{h}^{(n)}_{\mu}+ d_{\mu}^{(n)},
\label{mv-sep-3}
\end{equation}
\begin{equation}
h^{(n)}_{\mu}=-\frac{1}{2g}\epsilon_{\mu \nu \rho}\,
\hat{n}\cdot(\partial_\nu\hat{n}\times \partial_\rho\hat{n})
\makebox[.5in]{,}
\tilde{h}^{(n)}_{\mu}=\epsilon_{\mu \nu \rho}\partial_\mu C^{(n)}_\rho.
\end{equation}
\vspace{.1in}

For a monopole/anti-monopole correlated with a pair of center vortices, the
terms representing the defects in eqs. (\ref{mv-sep-3}) and (\ref{mv-sep}) are
given by \cite{lucho},
\begin{equation}
d_{\mu}^{(n)}=d_{\mu}^{(1)}+d_{\mu}^{(2)}
\makebox[.5in]{,}
d_{\mu\nu}^{(n)}=d_{\mu \nu}^{(1)}+d_{\mu \nu}^{(2)},
\end{equation}
\begin{equation}
d_\mu^{(\alpha)} =\frac{2\pi}{g} \int d\sigma\,
\frac{dx^{\alpha}_\mu}{d\sigma}\, \delta^{(3)}(x-x^\alpha(\sigma)).
\end{equation}
\begin{eqnarray}
d_{\mu \nu}^{(\alpha)}&=&\frac{2\pi}{g} \int d^2 \sigma_{\mu \nu}\,
\delta^{(4)}(x-x^\alpha(\sigma_1,\sigma_2)),
\label{den4}
\end{eqnarray}
Here, $x^\alpha(\sigma)$ (resp. $x^\alpha(\sigma_1,\sigma_2)$), $\alpha=1,2$, is
a pair of open center vortex
worldlines (worldsheets) with the same boundaries at $x^+$, $x^-$ (resp. $C^+$,
$C^-$), where the monopole and anti-monopole are localized. 
That is,
\begin{equation}
\partial_\mu d_\mu^{(\alpha)}
=\frac{2\pi}{g}(\delta^{(3)}(x-x^+)-\delta^{(3)}(x-x^-)),
\label{divg}
\end{equation}
\begin{equation}
\partial_\nu d_{\mu \nu}^{(\alpha)}=\frac{2\pi}{g} \left( \oint_{C^+} dy_\mu\,
\delta^{(4)}(x-y)- \oint_{C^-} dy_\mu\, \delta^{(4)}(x-y) \right).
\label{divf}
\end{equation}

For uncorrelated objects, we can write, $d_{\mu}^{(n)}=d_{\mu}^{(m)}
+d_{\mu}^{(v)}$, $d_{\mu\nu}^{(n)}= d_{\mu\nu}^{(m)} +d_{\mu\nu}^{(v)}$
\cite{lucho}, where the first part comes from defects in $\hat{n}_1$, 
$\hat{n}_2$ concentrated on open Dirac strings or worldsheets, while the second
part comes from defects localized on closed center vortex worldlines or
worldsheets, thus satisfying,
\begin{equation}
\partial_\mu d_\mu^{(v)}=0
\makebox[.5in]{,}
\partial_\nu d_{\mu \nu}^{(v)} =0.
\label{closed-div}
\end{equation}

\section{Wilson surfaces and frame defects}
\label{unc-c}

Up to now, we have seen how to represent the Wilson loop average in the
continuum, by considering an ensemble of thin defects.
In fact, in Yang-Mills theories, these defects are expected to be dressed by
quantum fluctuations, gaining dimensional properties such as the vortex
thickness and stiffness. This is the difficult part of the problem of
confinement, however, we can assume this scenario and analyze its feedback on
the structure of the theory. 

That is, we can replace the measure over the monopole and vortex sectors $[{\cal
D}U][{\cal D}V]$ by another one $[{\cal D}mon][{\cal D}vor]=[{\cal D}U][{\cal
D}V]\, e^{-S_d}$, including an action $S_d$ for the physical part of the
defects, characterizing the ensemble. The ensemble integration in eqs.
(\ref{WYMb}), (\ref{W3d}) can be separated to define an effective contribution
$S_{v,m}$,    
\begin{equation}
e^{-S_{v,m}[\bar{\lambda}_{\mu}]}=\int [{\cal D}{\rm mon}][{\cal D}{\rm vor}]\,
e^{i\frac{2\pi}{g}\sum \int dx_{\mu}\, \bar{\lambda}_{\mu}}
\makebox[.5in]{,}
\bar{\lambda}_{\mu}=\lambda_{\mu} + \frac{g}{2} s_{\mu},
\label{Svm3}
\end{equation}
\begin{equation}
e^{-S_{v,m}[\bar{\lambda}_{\mu \nu}]}=\int [{\cal D}{\rm mon}][{\cal D}{\rm
vor}]\, 
e^{i\frac{\pi}{g}\sum \int d^2 \sigma_{\mu \nu}\, \bar{\lambda}_{\mu \nu}}
\makebox[.5in]{,}
\bar{\lambda}_{\mu \nu}=\lambda_{\mu \nu} + g s_{\mu \nu}.
\label{Svm4}
\end{equation}
For correlated defects, with center vortices forming chains of monopoles and
anti-monopoles, the sum in the 
integrand would be performed over open center vortices attached in pairs to the
corresponding monopoles and anti-monopoles. In case of uncorrelated defects, the
sum would be over closed center vortices plus the sum over open Dirac strings
(in $3D$) or Dirac worldsheets (in $4D$).

It's still an open problem which ensemble is associated with $SU(2)$ Yang-Mills
theory. In the next section, we will discuss some possibilities in the framework
provided by the CFN decomposition and the PD representation in the presence of
defects.

Note that in the representation for $\bar{W}$, in eqs. (\ref{WYMb}) and
(\ref{W3d}), the terms containing $\epsilon_{\mu \nu \rho} \partial_\nu
s_{\rho}$, $\epsilon_{\mu \nu \rho \sigma} \partial_\nu s_{\rho \sigma} $,
according to eqs. (\ref{const-3d}) and (\ref{SC}), only depend on the Wilson
loop ${\cal C}$. However, because of the Wess-Zumino term in the PD
representation and the presence of defects, $\bar{W}$ contains a reference to
the initially considered $S({\cal C})$, although the usual nonabelian Wilson
loop representation contains no reference to a surface.

Terms in $d^{(n)}_{\mu}$, $d^{(n)}_{\mu \nu}$ associated with closed center
vortices, contribute with a flux $\pm 2\pi/g$ for each center vortex crossing
the surface. For a fixed Wilson loop ${\cal C}$, this contribution is
independent of the surface $S({\cal C})$ considered, given a factor
$(-1)^{link}$ that depends on the total linking number between the closed center
vortices and ${\cal C}$. When vortices percolate, this linking
gives an area law that displays N-ality \cite{greensite}.

As we have previously seen, monopoles can be joined by Dirac defects or by pairs
of open center vortices. 

In the first case, for a surface crossed by a Dirac defect the flux is $\pm 4\pi/g$, while for a surface that is not
crossed the flux is zero. Both situations contribute with a trivial phase $\pm 2\pi$, or zero,
respectively.

In the second case, consider for example a given monopole/anti-monopole configuration
joined by a pair of center vortices. If the loop ${\cal C}$ is ``linked"
by the chain, the flux contribution will be $+2\pi/g$ or $-2\pi/g$, depending on which center vortex in the pair
crosses the surface $S({\cal C})$. In both cases the Wilson loop gains a $-1$ factor.

However, we see that when considering the ensemble integration over defects, there are
singularities when the monopoles pass over $S({\cal C})$. 
This leads to the problem of how to obtain a representation of the Wilson loop
average with no reference to the initially considered Wilson surface $S({\cal C})$. The answer will 
depend on the type of ensemble. Initially we will discuss in the CFN-PD framework how, when the 
magnetic defects proliferate, the different phases can enable or preclude the possibility of performing large dual
transformations.  

\section{Possible ensembles and the associated closure properties of the dual
fields}
\label{sym-clos} 

As already discussed, the usual representation of the Wilson loop contains no reference to a surface, so that the
Petrov-Diakonov representation of the Wilson loop average should be invariant under the change
of initial Wilson surface $S({\cal C})$.

The consideration of a different $S({\cal C})$ can be written as the addition
of a closed surface $\partial \vartheta$, written as the border of a
three-volume $\vartheta$: $S({\cal C}) \rightarrow S({\cal C})\circ \, \partial
\vartheta$. This change can also be written in terms of the new sources, $s_\mu+ \Delta s_\mu$, $s_{\mu
\nu}+\Delta s_{\mu \nu}$ where, as $\partial \vartheta$ has no border, the
additional pieces verify,
\begin{equation}
\epsilon_{\mu \nu \rho} \partial_\nu \Delta s_\rho=0
\makebox[.5in]{,}
\epsilon_{\mu \nu \rho \sigma} \partial_\nu \Delta s_{\rho \sigma} =0 ,
\label{csurface}
\end{equation}
so that in $3D$ and $4D$ we can write, 
\begin{equation}
\frac{g}{2}\, \Delta s_\mu=\partial_\mu \omega^{(3)}
\makebox[.5in]{,}
g\,\Delta s_{\mu \nu}=\partial_\mu \omega^{(4)}_\nu -\partial_\nu
\omega^{(4)}_\mu .
\label{sclosed}
\end{equation}
Note that as long as $x$ is not on the closed surface $\partial \vartheta$, we
have $\partial_\mu \omega^{(3)}=0$, $\partial_\mu \omega^{(4)}_\nu -\partial_\nu
\omega^{(4)}_\mu=0$. That is, $\omega^{(3)}(x)$ is piecewise constant. It takes
the value $\pm g/2$, when $x$ is inside $\vartheta$, and is zero outside. The
plus or minus sign depends on whether the normal to $\partial \vartheta$ has an
internal or external orientation. 

In $4D$, the solution to eq. (\ref{sclosed}) is $\omega^{(4)}_\mu=\partial_\mu
\omega^{(4)}$, where $\omega^{(4)}$ is a multivalued phase. That is, when a path
linking the surface $\partial \vartheta$ is followed, $\omega^{(4)}$ changes by
an amount $\pm g/2$, while it does not change otherwise.

Now it is obvious that for a given $\lambda_\mu$, $\lambda_{\mu \nu}$ in the
integrand of eqs. (\ref{WYMb}), (\ref{W3d}), the configurations,
\begin{equation}
\lambda_\mu + \partial_\mu \omega
\makebox[.5in]{,}
\lambda_{\mu \nu} + \partial_\mu \omega_\nu - \partial_\nu \omega_\mu ,
\label{comega}
\end{equation}
with $\omega$ and $\omega_\mu$ smooth well-defined fields, always correspond to
another possible field configuration, so that we can operate with the associated
changes of variables as usual. Then
we are tempted to always consider,
\begin{equation}
\lambda_\mu \rightarrow \lambda_\mu + \partial_\mu \omega^{(3)}
\makebox[.5in]{,}
\lambda_{\mu \nu} \rightarrow \lambda_{\mu \nu} + \partial_\mu \omega^{(4)}_\nu
- \partial_\nu \omega^{(4)}_\mu 
\label{omega-trans}
\end{equation}
as an acceptable change of variables. In terms of the Hodge decomposition,
\begin{equation}
\lambda_{\mu}=\partial_\mu \phi+ B_{\mu}
\makebox[.5in]{,}
\lambda_{\mu \nu}=\partial_\mu \phi_\nu-\partial_\nu \phi_\mu +B_{\mu \nu},
\label{lambda}
\end{equation}
\begin{equation}
\partial_\mu B_{\mu}=0
\makebox[.5in]{,}
\partial_\mu \phi_\mu=0  
\makebox[.5in]{,}  \partial_\nu B_{\mu \nu}=0,
\label{g-fixing}
\end{equation}
we are asking about the possibility of considering changes of variables, 
\begin{equation}
\phi \rightarrow \phi + \omega^{(3)}
\makebox[.5in]{,}
\phi_\mu \rightarrow \phi_\mu + \omega^{(4)}_\mu = \phi_\mu + \partial_\mu
\omega^{(4)} .
\label{subs-phi}
\end{equation}
As we will see, this is not always possible and will depend on how the
symmetries are realized in the effective 
description for the Yang-Mills theory. In the next subsections we will discuss
some effective models; to simplify, we will consider the partition functions,
obtained by setting the sources $s_\mu$, $s_{\mu \nu}$ equal to zero in eqs.
(\ref{WYMb}), (\ref{W3d}).

\subsection{Correlated monopoles and center vortices in $3D$}
\label{corr3}

Center vortices have been discussed in the $SU(N)$ Georgi-Glashow model in $3D$
\cite{3}. Classically, this model contains vortices with topological charge
$Z(N)$. At the
quantum level, the vortex sector can be represented by means of vortex operators
associated with the monopole singularities in Euclidean spacetime,
where the vortices are created or destroyed. The relevant Green's functions are
incorporated  by means of 
an effective lagrangian for a vortex field,
\begin{equation}
\partial_\mu \bar{V}\partial_\mu V +\mu^2 \bar{V}V +\alpha (\bar{V}V)^2 +\beta 
(V^N +\bar{V}^N),
\label{vcond}
\end{equation}
which displays a global $Z(N)$ symmetry. When the vortex is an elementary
excitation ($\mu^2> 0$), there is no SSB. If vortices condense, SSB occurs ($\mu^2< 0$) and the formation
of a domain wall between a heavy quark-antiquark pair leads to an area law for
the Wilson loop \cite{3}.

Let us discuss the relationship between our representation and the effective model in eq. (\ref{vcond}). 
In the phase where the vortex is an elementary excitation with mass $\mu$, center vortex worldlines
can be associated with the propagation of point-like particles. Because 
of the coupling $e^{i\frac{2\pi}{g}\sum \int dx_{\mu}\, \lambda_\mu}$, when representing
this ensemble of worldlines in terms of an effective complex field $V(x)$, the vector field $\lambda_\mu$ in $S_{v,m}[\lambda_{\mu}]$ (cf. eq. (\ref{Svm3})) should be coupled through the covariant derivative,
\[
D_\mu V = [\partial_\mu +i (2\pi/g) \lambda_\mu ]\, V .
\]

In order to determine the possible terms in $S_{v,m}[\lambda_{\mu}]$, let us consider a transformation 
$\lambda_{\mu} \rightarrow \lambda_{\mu}+ \partial_\mu \omega$, with smooth $\omega$. 
In this case, the integrand in eq. (\ref{Svm3}) would gain a nontrivial factor,
\begin{equation} 
e^{i\frac{2\pi}{g}\sum \int dx_{\mu}\, \partial_\mu \omega}=e^{i\frac{4\pi}{g}\sum (\omega(x^+_i)-\omega(x^-_j))}.
\label{omega-change}
\end{equation}
Here, we used that center vortices are always attached in pairs to monopoles (anti-monopoles) located at $x^+_i$ ($x^-_j$). 
Therefore, when center vortices concatenate monopoles to form closed chains, we see that the presence of the 
monopoles should lead to an explicit $\omega$-symmetry breaking in $S_{v,m}$. On the other hand, the
possible terms in $S_{v,m}$ must be constrained by a symmetry, that in the phase where $\mu^2>0$ is expected to be displayed by the vacuum of the theory. When performing the $\omega^{(3)}$-transformation in eq. (\ref{omega-trans}),
the associated factor in eq. (\ref{omega-change}) is $e^{i\frac{4\pi}{g}\sum (\omega^{(3)}(x^+_i)-\omega^{(3)}(x^-_j))}=
e^{\pm i\frac{4\pi}{g} (N_+-N_-)\frac{g}{2}}=1$, where $N_+$ ($N_-$) is the number of monopoles (anti-mon\-o\-poles) in $\vartheta$.   

Therefore, the natural result for the ensemble integration over chains
is of the form,
\begin{equation}
S_{v,m} = \overline{D_\mu V}D_\mu V + \mu^2 \bar{V}V +\alpha (\bar{V}V)^2 +\beta
(V^2 +\bar{V}^2)+ 
S_0 [\tilde{F}_\mu],
\label{ensemble-int}
\end{equation}
where $\tilde{F}_\mu=\epsilon_{\mu \nu \rho}\partial_\nu \lambda_{\rho}$. 
This $S_{v,m}$ enjoys the desired properties, as the $\omega$-symmetry is explicitly broken by the $V^2$, $\bar{V}^2$ terms. In
addition, it displays a local $Z(2)$ symmetry $V \to e^{-i\frac{2\pi}{g}\omega^{(3)}}V$, 
$\lambda_{\mu} \rightarrow \lambda_{\mu}+ \partial_\mu \omega^{(3)}$. This comes about as $\omega^{(3)}$ is given by $\pm g/2$ inside $\vartheta$, while it is zero outside. Then, this transformation changes the sign of $V$, $\bar{V}$ inside $\vartheta$, thus leaving the $V^2$, $\bar{V}^2$ terms invariant. 
The term $S_0$ is also invariant; this can be seen from the property $\epsilon_{\mu \nu \rho}\partial_\nu \partial_{\rho}\omega^{(3)}=0$, implied from eqs. (\ref{csurface}), (\ref{sclosed}). For a discussion of local 
discrete transformations in $3D$ gauge theories, when matter fields in the
fundamental representation are present, see refs. \cite{cesar,1.7}.

The effective contribution in eq. (\ref{ensemble-int}) can be also obtained by direct ensemble integration 
based on polymer field theory techniques, considering a phase where center vortices are flexible, characterized by a small stiffness, and
tensile, weighted by a factor $e^{-\mu L}$ \cite{AO}. 

Then, taking into account the other terms in eq. (\ref{W3d}) and the integral
over $[{\cal D}\Psi]$, the effective model for the partition function in $SU(2)$
Yang-Mills, including the effect of chains, would be of the form (for a discussion of the $[{\cal D}\Psi]$ integration,
see ref. \cite{lucho} and references therein),
\begin{equation}
S_{{\rm eff}} = \overline{D_\mu V}D_\mu V + \mu^2 \bar{V}V +\alpha (\bar{V}V)^2
+\beta (V^2 +\bar{V}^2)+ 
S [\tilde{F}_\mu] + \gamma\lambda_\mu \lambda_\mu ,
\label{mod}
\end{equation}
where the term $\lambda_\mu \lambda_\mu$ explicitly breaks the $\omega^{(3)}$-symmetry in $S_{{\rm eff}}$, preserving
a global $Z(2)$. Now, in a phase where this global $Z(2)$ symmetry is spontaneously broken ($\mu^2<0$), 
we have a topological structure, whose existence depends on the consideration of
well-behaved continuous fields. In particular, we will have finite action domain
walls where $V(x)$ will continuously change from $+V_0$ to $-V_0$, accompanied
by a well-behaved continuous $\lambda_\mu$. As we go across the wall, the phase
of $V(x)$ must either change continuously from $0$ to $\pi$, or we can have a
$\pi$ discontinuity at a thin surface $S$ inside the thick wall, as long as
$V(x)$ vanishes for points $x \in S$. These kinds of walls have been discussed
in ref. \cite{cesar}. 

In other words, when the global $Z(2)$ is spontaneously broken, changes of
variables of the form 
$V \to e^{-i\frac{2\pi}{g}\omega^{(3)}}V$, or $\lambda_\mu \rightarrow \lambda_\mu +
\partial_\mu \omega^{(3)}$ are not acceptable, as the fields produced will no
longer correspond to well-behaved continuous fields. On the other hand, in the
phase where the global $Z(2)$ symmetry is not spontaneously broken ($\mu^2>0$), 
these requirements are no longer applicable,
and the large dual transformations are acceptable.

It is also interesting to note that if $S[\tilde{F}_\mu]$ were dominated
by a Maxwell term (see ref. \cite{lucho}), $\lambda_\mu$ would be a massive
vector field. Then, depending on the generated mass scale, $\lambda_\mu$ would be
suppressed and the model in eq. (\ref{vcond}) would be obtained. In addition, because of eq.
(\ref{const-3d}), the off-diagonal current is given by $\epsilon_{\mu \nu
\rho}\partial_\nu \lambda_\rho$ (in this subsection we are considering
$s_\mu=0$) so that this suppression would correspond to Abelian dominance
\cite{ad-latt,ad}.

\subsection{Loop-like monopoles in $4D$}

In $4D$, the problem concerning the closure properties of large dual field
transformations can be easily understood
in the simpler context of ensembles of uncorrelated monopoles and center
vortices. In this case, the ensemble integration is of the form
$S_{v,m}[\lambda_{\mu \nu}]=S_v[B_{\mu \nu}]+S_m[\phi_\mu]$,
\begin{equation}
e^{-S_v[B_{\mu \nu}]}=\int[{\cal D}{\rm vor}]\, e^{i\frac{\pi}{g}\sum_{v} \oint
d^2 \sigma_{\mu \nu}\, B_{\mu \nu}},
\label{vort-act}
\end{equation}
\begin{equation}
e^{-S_m[\phi_{\mu}]}=\int[{\cal D}{\rm mon}]\, e^{i\frac{4\pi}{g}\sum_{ij}
\left( \oint_{C^+_j} dy_\mu\, \phi_\mu- \oint_{C^-_i} dy_\mu\, \phi_\mu
\right)},
\end{equation}
where we have used that unobservable Dirac worldsheets can be decoupled in favor
of their borders (see ref. \cite{AML}). 

As the dual vector field $\phi_\mu$ is minimally coupled with closed string-like
objects, the action $S_{v,m}$ originated from the ensemble integration will be
gauge invariant under regular gauge transformations $\phi_\mu \rightarrow
\phi_\mu + \partial_\mu \omega$, and will contain a complex field $\Phi$
representing the monopole sector minimally coupled through the covariant
derivative (for a review, see ref. \cite{antonov}), 
\[
[ \partial_\mu +i (4\pi/g) \phi_\mu ]\, \Phi .
\]
Now, in the corresponding effective action for $SU(2)$ Yang-Mills, the
$\lambda_{\mu \nu}\lambda_{\mu \nu}$ term and the ${\cal D}\Psi$ integration in
eq. (\ref{WYMb}) will give additional gauge invariant terms, depending on
$\partial_\mu \phi_\nu - \partial_\mu \phi_\nu$. 

In a phase where the $U(1)$ gauge symmetry is spontaneously broken, we will
again have a topological structure, whose existence  depends on the
consideration of well-behaved continuous fields. For instance, the phase in
$\Phi(x)$ can be ill-defined only in places of false vacuum. Therefore, when SSB
is present, changes of variables with multivalued phase $\omega^{(4)}$ cannot be
accepted, as in general $e^{-i\frac{4\pi}{g}\omega^{(4)}}\Phi$ would be ill-defined on the
closed surface $\partial \vartheta$. 

This discussion, together with the minimal coupling with $\phi_\mu$, leads to
the impossibility of considering $\phi_\mu \rightarrow \phi_\mu + \partial_\mu
\omega^{(4)}$ as an acceptable change of variables in the path-integral for a
SSB phase. A similar situation occurs with the spacetime independent phase
transformations, in the SSB phase, where the boundary condition imposed on
$\Phi$ at infinity is not closed under them. 

In more formal language, according to the Elitzur theorem \cite{Elitzur}, gauge
transformations cannot be spontaneously broken. That is, at the nonperturbative
level, in the canonical version of the quantized theory, there is no gauge variant 
operator with a nonzero expectation value (for
a discussion in the context of confinement, see refs. \cite{GL1,CG1}). 

What can be spontaneously broken is the subgroup of ``global'' gauge
transformations that remains after a gauge fixing is implemented. An order
parameter to explore the possible realizations must be something invariant under
gauge transformations and variant under global transformations. This can be
constructed for different gauge fixings. In the dual $\hat{\phi}_\mu$-theory it
could be considered of the form,
\begin{equation}
\hat{O}=e^{i\frac{4\pi}{g} \int d^4 x'\, \partial_\mu \hat{\phi}_\mu (x') D(x'-x)}\,
\hat{\Phi}(x),
\label{opam} 
\end{equation}
where $D(x)$ is the Green function for the Laplacian operator. This order
parameter is invariant under local regular phase transformations $\hat{\phi}_\mu
\rightarrow \hat{\phi}_\mu + \partial_\mu \alpha(x)$, $\hat{\Phi} \to
e^{-i\frac{4\pi}{g} \alpha(x)} \hat{\Phi}$, while under spacetime independent ones it transforms
as, $\hat{O} \rightarrow e^{i\alpha} \hat{O}$.

We also note that $\omega^{(4)}$ satisfies $\partial_\mu \partial_\mu
\omega^{(4)} =0$ (see subsection \S \ref{TI}), so that the order parameter in
eq. (\ref{opam}) also transforms under the operation, $\hat{\phi}_\mu
\rightarrow \hat{\phi}_\mu + \partial_\mu \omega^{(4)}$, $\hat{\Phi} \to
e^{-i\frac{4\pi}{g}\omega^{(4)}}\hat{\Phi}$, according to $\hat{O} \rightarrow
e^{-i\frac{4\pi}{g}\omega^{(4)}} \hat{O}$.

Then, when the spacetime independent phase transformations are spontaneously
broken, the large dual transformations are also spontaneously broken, that is,
the vacuum is not invariant under them.

\subsection{Correlated monopoles and center vortices in $4D$}

For chains of monopoles and anti-monopoles, we have (cf. eqs. (\ref{den4}),
(\ref{divf})),
\begin{equation}
e^{-S_{v,m}[B_{\mu \nu}, \phi_{\mu}]}=\int[{\cal D}{\rm vor}][{\cal D}{\rm
mon}]\, e^{i\frac{\pi}{g}\sum_{v} \int d^2 \sigma_{\mu \nu}\, B_{\mu
\nu}+i\frac{4\pi}{g}\sum_{ij} \left( \oint_{C^+_j} dy_\mu\, \phi_\mu-
\oint_{C^-_i} dy_\mu\, \phi_\mu \right)},
\label{vort-act-chains}
\end{equation}
where $B_{\mu \nu}$ is integrated over open vortex worldsheets with their
borders attached in pairs, so as to form  the associated monopole or
anti-monopole loops at $C^+_j$, $C^-_i$. 

For each vortex worldsheet, we have a contribution in the integrand of the
form, 
\begin{equation}
V(C^+) V(C^-)\, e^{i\frac{\pi}{g} \int_{\Sigma} d^2 \sigma_{\mu \nu}\, B_{\mu
\nu}}
\makebox[.5in]{,}
V(C^{\pm})=e^{\pm i \frac{2\pi}{g} \oint_{C^{\pm}} dy_\mu\, \phi_\mu},
\end{equation}
where $\Sigma=\Sigma (C^+, C^-)$ is a surface with borders at $C^+$ and $C^-$. 

This configuration represents the creation, propagation and annihilation of a
loop, minimally coupled to $B_{\mu \nu}$, so that $V(C)$ can be compared to the
disorder operator introduced in ref. \cite{3} for the Yang-Mills theory.

If center vortex worldsheets were closed objects propagating string-like excitations,
characterized by a finite tension, $S_{v,m}[\lambda_{\mu \nu}]$
in eq. (\ref{Svm4}) would be invariant under the transformations $\lambda_{\mu
\nu} \rightarrow \lambda_{\mu \nu}+ \partial_\mu \omega_\nu -\partial_\nu
\omega_\mu$, including  the large ones, $\phi_\mu \rightarrow \phi_\mu
+\partial_\mu \omega^{(4)}$, so that a typical effective action for this sector
would be of the form \cite{ambjorn}-\cite{franz},
\begin{equation}
S_{c.v.} = S_{c.v.}[\tilde{H}_\mu]
\makebox[.5in]{,}
\tilde{H}_\mu=\epsilon_{\mu \nu \rho \sigma}\partial_\nu \lambda_{\rho \sigma}
\end{equation}
(note that due to eqs. (\ref{csurface}), (\ref{sclosed}), 
$\epsilon_{\mu \nu \rho \sigma} \partial_\nu \partial_{\rho}\partial_\sigma \omega^{(4)}=0$).
When these center vortex worldsheets concatenate monopoles, we can see from eq. (\ref{vort-act-chains}) that 
the presence of the latter explicitly breaks the $\omega_\mu$-symmetry
in $S_{v,m}[\lambda_{\mu \nu}]$. However, this
contribution will be symmetric under the regular $\phi_\mu \rightarrow \phi_\mu
+\partial_\mu \omega$ transformations and, as in eq. (\ref{vort-act-chains}) the
loop variables appear in the form $V^2(C^{\pm})$, it is expected to be
symmetric under the large ones, $\phi_\mu \rightarrow \phi_\mu +\partial_\mu
\omega^{(4)}$. Then, $S_{v,m}$ can be written as $S^{(4)}+S_0[\tilde{H}_\mu]$,
where $S^{(4)}$ is only symmetric under $\omega^{(4)}$-transformations, breaking the $\omega_\mu$-symmetry. 
This part would be analogous to the $V$-dependent terms in eq. (\ref{ensemble-int}), 
however, the problem of presenting effective models for $S^{(4)}$ is highly nontrivial, 
as in $4D$ the vortex field $V(x)$ is replaced by a loop variable $V(C)$.

Taking into account the other terms in the representation
(\ref{WYMb}) and the $[{\cal D}\Psi]$ integration \cite{lucho}, in this case,
the effective action for Yang-Mills is expected to be of the form,
\begin{equation}
S_{{\rm eff}} = S^{(4)}+ S [\tilde{H}_\mu] + \gamma
\lambda_{\mu \nu} \lambda_{\mu \nu} ,
\end{equation}
where the $\lambda_{\mu \nu} \lambda_{\mu \nu}$ term explicitely breaks the $\omega^{(4)}$-symmetry present in the first two terms. 
Again, we could expect a phase for the ensemble of chains where the associated regularity requirements imposed on $\lambda_{\mu \nu}$ could disallow the changes of variables $\phi_\mu \rightarrow \phi_\mu + \partial_\mu
\omega^{(4)}$, as occurs in $3D$ with the $\mu^2<0$ phase and the changes of
variables $\phi \rightarrow \phi + \omega^{(3)}$ (see subsection \S \ref{corr3}).

\section{Wilson surface decoupling vs. Wilson surface variables}
\label{dec-obs}

The discussion about how a surface whose border is the Wilson loop can become
observable in Yang-Mills theory is a key point in understanding the possible
mechanisms underlying confinement and its associated properties. 

In ref. \cite{3}, the possible observability of Wilson surfaces or center vortex
worldsheets has been analyzed as follows. The algebra between the Wilson loop
operator $\hat{W}({\cal C},t)$ and the disorder operator 
$\hat{V}(C',t)$ is,
\begin{equation}
\hat{W}({\cal C},t)\hat{V}(C',t) = \hat{V}(C',t) \hat{W}({\cal C},t)
(-1)^{link},
\label{conmu}
\end{equation}
where $C$ and $C'$ are defined at a given time $t$, and the right-hand side
contains the linking number between them. Then, a family $C'(a)$ in $R^4$, $a\in
[0,1]$ is considered, continuously changing from $C'_0$, passing by an
intermediate $C'_i$, and then back to $C'_0$, both curves living on the constant
time $t$ hyperplane where $C$ is contained. As we are in $R^4$, this family can
be chosen with $C'_0$ (resp. $C'_i$) unlinked (resp. linked) with $C$, and such
that $C'(a)$ never comes close to $C$. In these conditions, a declustering
property was used,
\begin{equation}
\langle W({\cal C}) V(C'_a) \rangle \approx \langle W({\cal C})\rangle \langle
V(C'_a) \rangle \,
e^{i\alpha ({\cal C},C'_a)},
\end{equation}
where the phase is required in order to be consistent with eq. (\ref{conmu}),
which implies that $e^{i\alpha ({\cal C},C'_a)}$ must change from $+1$ to $-1$
and then back to $+1$ in this process. If massless modes exist in Yang-Mills,
$\alpha ({\cal C},C'_a)$ could be a smoothly varying function. On the other
hand, when $Z(2)$-invariant Higgs fields are switched on, it has been argued
that a sudden
change in the phase must exist, and as the pairs of curves are always mantained
far apart, an observable surface must
be attached to the Wilson loop or to the half-charge magnetic loop.

In ref. \cite{polyakov1}, the Wilson loop average $\bar{W}$ has been analyzed in
confining models such as compact $QED(3)$ and $QED(4)$, the latter regularized
on the lattice. In that reference, considering the dual field $\phi$ defined on
the interval $[-\infty, +\infty]$, a representation based on axion fields, with
multivalued action, has been obtained, and a series of approximations led to an
explicit dependence of the resulting $\bar{W}$ on the arbitrary $S({\cal C})$
appearing in its definition. Then it has been conjectured that this problem
would be resolved if all the branches of the multivalued action were considered
in the calculation, and that this would be equivalent to considering the
integration over all Wilson surfaces (that now become dynamical) and dual
$\tilde{\phi}$'s with an appropriate jump at the associated surface. 

Because of the Wess-Zumino term in the PD representation, our expressions in
eqs. (\ref{WYMb}), (\ref{W3d}) for  $\bar{W}$ in Yang-Mills theory also have an
arbitrary surface $S({\cal C})$ attached to the Wilson loop from the beginning.
However, the representation must be independent of $S({\cal C})$. In the next
subsections, we will discuss how to obtain, in general, a Wilson loop
representation with no reference to the initially considered $S({\cal C})$. 

The answer will depend on the underlying realization of symmetries in the
effective models describing the Yang-Mills theory, which according to the
discussion in section \S \ref{sym-clos} will determine whether changes of
variables $\phi \rightarrow \phi + \omega^{(3)}$, $\phi_\mu \rightarrow \phi_\mu
+ \partial_\mu \omega^{(4)}$ are acceptable or not. 

In $3D$, we have seen that in the phase without global $Z(2)$ SSB, the change
$\phi \rightarrow \phi + \omega^{(3)}$ is acceptable; this corresponds to
single-valued possibly discontinuous $\phi$'s defined on the interval $[-\infty,
+\infty]$. In this case, we will be able to decouple the initial Wilson surface
following treatment I. On the other hand, in the SSB phase, the change is not
acceptable, and the $\phi$'s will have to be considered as continuous
multivalued angles. Here, the reference to the arbitrary initial $S({\cal C})$
will also dissapear, but giving place to an integral over all the Wilson
surfaces, and the above mentioned  $\tilde{\phi}$'s. These two possibilities for
the class of $\phi$'s and their consequences, will also be extended to classes
of $\phi_\mu$'s in $4D$ theories in the continuum.
 
\subsection{Dealing with Wilson surfaces I}
\label{TI}

Let us consider $\phi$, $\phi_\mu$ as single-valued fields, so that the large
dual transformations, adding the single-valued pieces $\omega^{(3)}$,
$\partial_\mu \omega^{(4)}$, can be performed. Of course, in this case, the
Wilson surface should be an unobservable object, but the question is, how can we
use the large dual transformations in order to  decouple $S({\cal C})$ in favor
of ${\cal C}$, thus evidencing the unobservability of $S({\cal C})$.

For this aim, let us follow a procedure similar to the one we implemented in
ref. \cite{AML}, where we discussed how to decouple unobservable Dirac defects
in favor of their borders, in the CFN representation of the Yang-Mills partition
function.

Considering the auxiliary fields $\zeta_\mu$, $\zeta_{\mu \nu}$, and a change of
variables $\lambda_\mu+\frac{g}{2}\, s_{\mu}\rightarrow \lambda_\mu$,
$\lambda_{\mu \nu}+ g s_{\mu \nu}\rightarrow \lambda_{\mu \nu}$, we have,
\begin{eqnarray}
\bar{W}({\cal C}) &=&  \int [{\cal D}\zeta][{\cal D}\lambda][{\cal D}\Psi]\,
e^{-S_c-\int d^3x\, \frac{1}{2}\zeta_\mu \zeta_\mu}\times\nonumber \\
&&\times e^{i\int d^3x\, \{(\lambda_{\mu}-\frac{g}{2}\, s_{\mu})(\zeta_\mu
+k_{\mu})+(\epsilon_{\mu \nu \rho} \partial_\nu \lambda_{\rho}-
J^c_\mu ) A^{(n)}_\mu +\lambda_{\mu } d^{(n)}_{\mu}\}},
\label{Wzeta1}
\end{eqnarray}
\begin{eqnarray}
\bar{W}({\cal C})
&= &\int [{\cal D}\zeta][{\cal D}\lambda][{\cal D}\Psi]\, e^{-S_c-\int d^4x\,
\frac{1}{4}\zeta_{\mu \nu} \zeta_{\mu \nu}}\times\nonumber \\
&&\times e^{i\int d^4x\, \{ \frac{1}{2}(\lambda_{\mu \nu}-g s_{\mu \nu})
(\zeta_{\mu \nu}+ k_{\mu \nu})+(\frac{1}{2}\epsilon_{\mu \nu \rho \sigma}
\partial_\nu \lambda_{\rho \sigma}-
J^c_\mu ) A^{(n)}_\mu +\frac{1}{2}\lambda_{\mu \nu}d^{(n)}_{\mu \nu}\}
}.\nonumber \\
\label{Wzeta2}
\end{eqnarray}
The path-integrals in $[{\cal D} \lambda]$ can be done over the fields defined
in eq. (\ref{lambda}), with $\phi$, $\phi_\mu$ single-valued. Including the
conditions in eq. (\ref{g-fixing}), in $3D$ we must consider the replacement,
\begin{equation}
[{\cal D}\lambda]\rightarrow [{\cal D}B][{\cal D}\phi][{\cal D}\xi]\, e^{i\int
d^4x\, \xi\, \partial_\mu B_{\mu}},
\end{equation}
while in $4D$, we have,
\begin{equation}
[{\cal D}\lambda]\rightarrow [{\cal D}B][{\cal D}\phi][{\cal D}\xi][{\cal
D}\gamma]\, 
e^{i\int d^4x\, \xi_\mu \partial_\nu B_{\mu \nu}} e^{i\int d^4x\, \gamma\,
\partial_\mu \phi_\mu}.
\label{gfdual}
\end{equation}

Therefore, using eq. (\ref{sclosed}) and considering in eqs. (\ref{Wzeta1}) and
(\ref{Wzeta2}) the large dual transformations, with trivial Jacobian,
\begin{equation}
\phi \rightarrow \phi - \omega^{(3)} 
\makebox[.5in]{,}
\phi_{\mu} \rightarrow \phi_{\mu} -\partial_\mu \omega^{(4)} ,
\label{change34}
\end{equation}
the terms in eqs. ({\ref{Wzeta1}}), (\ref{Wzeta2}) containing respectively
$d^{(n)}_{\mu}$, $d^{(n)}_{\mu \nu}$, gain a phase which is a trivial multiple
of $2\pi$, the second term is invariant, while the first term gives a change in
the surface. In the $4D$ case, it is important to underline that the explicit
form for $\partial_\mu \omega^{(4)}$ is,
\begin{equation}
\partial_\mu \omega^{(4)} =\pm \frac{g}{2}\int_{\vartheta} d^{3} \tilde{\sigma
}_{\nu }\,
(\delta_{\mu \nu } \partial^{2} - \partial_{\mu } \partial_{\nu } )
D(x-\bar{x} (\sigma )),
\label{dechi}
\end{equation} 
\begin{equation}
d^{3} \tilde{\sigma }_{\mu}=\frac{1}{2} \epsilon_{\mu \alpha \beta \gamma}\,
\epsilon_{ijk}\frac{\partial \bar{x}_{\alpha}}{\partial
\sigma_{i}}\frac{\partial \bar{x}_{\beta}}{\partial \sigma_{j}} \frac{\partial
\bar{x}_{\gamma}}{\partial \sigma_{k}}\, d\sigma_1 d\sigma_2 d\sigma_{3}.
\end{equation}
Using Stokes' theorem, this can be written only in terms of $\partial
\vartheta$, the manifold where the added closed Wilson surface is placed (for a
discussion in the context of thin center vortices and Dirac worldsheets, see
refs. \cite{engelhardt1,reinhardt,AML}). Therefore, the index structure in eq.
(\ref{dechi}) implies $\partial_\mu \partial_\mu \omega^{(4)} =0$, and
$\partial_\mu \phi_\mu$ in the measure given in eq. (\ref{gfdual}) is invariant
under the change of variables in eq. (\ref{change34}). 

Summarizing, in $3D$ and $4D$ we can deform the Wilson surface by means of a
change of variables, with trivial Jacobian, keeping its border ${\cal C}$ fixed.

Now let us consider a Hodge decomposition,
\begin{equation}
\zeta_{\mu}+ k_\mu=\partial_\mu \psi + C_{\mu}
\makebox[.5in]{,}
\zeta_{\mu \nu} + k_{\mu \nu}=\partial_\mu \psi_\nu-\partial_\nu \psi_\mu
+C_{\mu \nu},
\label{zeta}
\end{equation}
with,
\begin{equation}
\partial_\mu C_{\mu}=0
\makebox[.5in]{,}
\partial_\nu C_{\mu \nu}=0  
\makebox[.5in]{,}  
\partial_\mu \psi_\mu=0,
\end{equation}
that permits the identification of $C_\mu$, $C_{\mu \nu}$ as fields only coupled
to the Wilson loop ${\cal C}$,  while the fields $\psi$, $\psi_\mu$ are the ones
coupled with the whole surface $S({\cal C})$.

We will show that the Wilson surface can be decoupled by means of an appropriate
change of variables, leaving only the effect of its border. For this purpose we
leave the integration over $\zeta_\mu$, $\zeta_{\mu \nu}$ and the charged fields
present in $k_\mu$, $k_{\mu \nu}$ until the end, and analyze the integral over
$\lambda_\mu$, $\lambda_{\mu \nu}$ first. Let us consider the term coupling
$\psi$, $\psi_\mu$, 
\begin{equation}
J_{S({\cal C})}=\left\{ \begin{array}{ll}
\int d^3x\, s_\mu \partial_\mu \psi & \\
\int d^4x\, s_{\mu \nu} (\partial_\mu \psi_\nu-\partial_\nu \psi_\mu).& 
\end{array}\right.
\label{Scoupling}
\end{equation}
For the initial Wilson surface, and sources $s_\mu$, $s_{\mu \nu}$, we can
assume 
$J_{S({\cal C})}>0$ without loss of generality. In addition, we can assume that
a closed surface $\partial \vartheta$ exists, such that,
\begin{equation}
J_{[\partial \vartheta]}=\left\{ \begin{array}{ll}
\int d^3x\, \Delta s_\mu \partial_\mu \psi  &  \\ 
\int d^4x\, \Delta s_{\mu \nu} (\partial_\mu \psi_\nu-\partial_\nu \psi_\mu),& 
\end{array}\right. 
\end{equation}
is nonzero. In this regard, it suffices to consider a small $\vartheta$, as in
this case $J_{[\partial \vartheta]}$ is given by the local value of
$\partial^2\psi$, $\partial^2\psi_\mu$. If this value were zero for any
$\partial \vartheta$, we would have $\psi \equiv 0$, $\psi_\mu\equiv 0$, and the
term coupling the surface would be automatically zero. 

Now let us include $m$ times the closed surface $\partial \vartheta$ and define
the sources $s'_{\mu}$, $s'_{\mu \nu}$, concentrated on the surface $S'({\cal
S})=S({\cal C})\circ \, [\partial \vartheta]^m$. This amounts to the
transformation,
\begin{equation}
\phi \rightarrow \phi - m\omega^{(3)} 
\makebox[.5in]{,}
\phi_{\mu} \rightarrow \phi_{\mu} -\partial_\mu (m\omega^{(4)}) .
\end{equation}
Then, we have,
\begin{equation}
J_{S'({\cal C})} = J_{S({\cal C})} + m J_{[\partial \vartheta]}.
\end{equation}
Now, we can take $\partial \vartheta$ oriented such that,
\begin{equation}
J_{[\partial \vartheta]} < 0 ,
\end{equation}
so that $J_{S'({\cal C})}$ can be rendered negative for a large enough value of
$m$. As, $S'({\cal C})$ can be continuously deformed into $S({\cal C})$, by
shrinking $\partial \vartheta$ to zero, an intermediate surface $S_0({\cal C})$
must exist in this process such that $J_{S_0({\cal C})}=0$ is verified. This
suggests that it is always possible to make a large dual transformation that
changes the initial $S({\cal C})$ into $S_0({\cal C})$ thus nullifying the terms
coupling the Wilson surface with $\psi$, $\psi_\mu$. Then, in practice, the
prescription in this case for obtaining a representation for $\bar{W}$ with no
reference to the initial $S({\cal} C)$ is simply to disregard the above
mentioned terms in eqs. (\ref{Wzeta1}) and (\ref{Wzeta2}).

\subsection{Dealing with Wilson surfaces II}
\label{TII}

Now the question is what to do in the case where the ensemble of defects
requires regular fields $\phi$, $\phi_\mu$ in the Hodge decomposition
(\ref{lambda}), so that large dual changes of variables are no longer
acceptable.  

In order to answer this question, let us first consider the $3D$ case, denoting
the fields in the decomposition for $\lambda_\mu$, with the properties used in
the previous subsection, as $\phi^{{\rm I}}$ and $B^{{\rm I}}_{\mu}$. That is,
$\phi^{{\rm I}}$ is single-valued, and defined on the interval $[-\infty,
+\infty]$. Now, considering a smooth $\lambda_\mu$, adding and substracting a
source $s_\mu(\tilde{\Sigma})$ concentrated on a general Wilson surface
$\tilde{\Sigma}$ whose border is ${\cal C}$, we can also write a decomposition
using fields $\phi^{{\rm II}}$ and 
$B^{{\rm II}}_{\mu}$, with $\phi^{{\rm II}}$ being a multivalued field, when we
go around the Wilson loop ${\cal C}$. That is,
\begin{equation}
\lambda_{\mu}=\partial_\mu \phi^{{\rm I}}+ B^{{\rm I}}_{\mu}=\partial_\mu
\phi^{{\rm II}}+ B^{{\rm II}}_{\mu},
\end{equation}
\begin{equation}
\partial_\mu \phi^{{\rm II}}=\partial_\mu \tilde{\phi}-\frac{g}{2}\,
s_\mu(\tilde{\Sigma})
\makebox[.5in]{,}
\tilde{\phi}= \phi^{{\rm I}}+\frac{g}{2}\, \partial^{-2} (\partial \cdot
s(\tilde{\Sigma})),
\end{equation}
\begin{equation}
B^{{\rm II}}_{\mu}= B^{{\rm I}}_{\mu}-\frac{g}{2}\, \epsilon_{\mu \nu
\rho}\partial_\nu
\partial^{-2} j_\rho ({\cal C}).
\end{equation}
Note that, when computing $\partial_\mu \phi^{{\rm II}}$, the derivative of the
discontinuity in the second term of $\tilde{\phi}$ is cancelled by the
$-\frac{g}{2}\, s_\mu(\tilde{\Sigma})$ term, so that the defined $\phi^{{\rm
II}}$ is a continuously changing multivalued field, satisfying $\epsilon_{\mu
\nu \rho} \partial_\nu \partial_{\rho}\phi^{{\rm II}} = -\frac{g}{2} j_\mu
({\cal C})$. 

On the other hand, the class of fields $\lambda_\mu$ generated by the
single-valued, possibly discontinuous,  $\phi^{{\rm I}}$'s is different from the
class of fields $\lambda_\mu$ generated by the continuous multivalued
$\phi^{{\rm II}}$'s. In the first case, there is no problem in summing
$\phi^{{\rm I}}$ and $\omega^{(3)}$ to obtain another possible configuration; in
the second case, summing the multivalued $\phi^{{\rm II}}$ and $\omega^{(3)}$
does not make any sense.  

In a similar way, in $4D$, we will have type I and type II dual fields
$\phi_\mu$, the former are the single-valued fields used in the previous
subsection, the latter being appropriate for describing situations where
$\omega^{(4)}$-changes of variables are not acceptable.

Then in this section, we will introduce a decomposition in terms of type II
fields in $3D$ and $4D$, enjoying the properties,
\begin{equation}
\epsilon_{\mu \nu \rho} \partial_\nu \partial_{\rho}\phi = -\frac{g}{2} j_\mu
({\cal C}),
\makebox[.5in]{,}
\epsilon_{\mu \nu \rho \sigma}\partial_\nu \partial_\rho \phi_\sigma
=-\frac{g}{2} j_\mu ({\cal C}).
\label{multi}
\end{equation}

In three dimensions, the integral of $\epsilon_{\mu \nu \rho} \partial_\nu
\partial_{\rho}\phi$ over an open surface with border ${\cal P}$, crossed by the
Wilson loop ${\cal C}$, gives $\pm g/2$. Then, using Stokes' theorem, the
integral of $\partial_\mu \phi$ along ${\cal P}$ gives $\Delta \phi=\pm g/2$,
while this change is zero on a path that does not link ${\cal C}$. We have
already seen that the multivalued $\phi$ can be written in terms of
 $\tilde{\phi}(x)$, discontinuous at some surface $\tilde{\Sigma}$ whose border
is the Wilson loop ${\cal C}$, such that $\partial_\mu \phi = \partial_\mu
\tilde{\phi} - \frac{g}{2} s_\mu(\tilde{\Sigma})$.

In four dimensions, $\phi_\mu$ must be considered as a vector field that cannot
be globally defined on the closed surfaces ${\cal S}$ linked by the Wilson loop.
It can be differently defined on two hemispheres meeting on a closed path ${\cal
P}$, where the difference between $\phi_\mu$ continued from each one of the
hemispheres is $\partial_\mu \alpha$, with $\alpha$ multivalued. This can be
visualized by considering, for example, the Wilson loop contained in the $x^0=0$
hyperplane (a three-volume). If we stay on this hyperplane, the loop ${\cal C}$
is seen to be linked by path ${\cal P}$. If we continuously move to other
hyperplanes with $x^0\neq 0$, the Wilson loop will no longer be seen, while the
former path ${\cal P}$ will be seen to continuously shrink to a point, mapping
both hemispheres in four dimensions, for positive or negative $x_0$, forming the
closed surface linked by ${\cal C}$.

Precisely because of eq. (\ref{multi}), the integral of $\epsilon_{\mu \nu \rho
\sigma}\partial_\nu \partial_\mu \phi_\nu$,  over an open three-volume with
border ${\cal S}$, gives $\pm g/2$ and can be equated via Gauss' theorem with
the integral of $\epsilon_{\mu \nu \rho \sigma}\partial_\rho \phi_\sigma$ over
the closed surface ${\cal S}$ linked by ${\cal C}$. This surface integral can be
done on the two hemispheres A and B, sharing the same border ${\cal P}$, where
$\phi_\mu$ takes the values $\phi^A_\mu$ and $\phi^B_\mu$, respectively. Now, we
can use Stokes' theorem to write the surface integral as the line integral of
$\phi^A_\mu -\phi^B_\mu=\partial_\mu \alpha$ over the closed path ${\cal P}$,
thus obtaining $\Delta \alpha =\pm g/2$. 

Then, in eq. (\ref{lambda}), the multivalued field $\phi_\mu$ can be replaced by
$\tilde{\phi}_\mu(x)$, defined on the whole Euclidean spacetime as a function of
point $x$ and discontinuous at some surface $\tilde{\Sigma}$, whose border is
the Wilson loop ${\cal C}$. Again, the derivatives of $\phi_\mu$ cannot contain
any singular term on $\tilde{\Sigma}$, so that the replacement must by done as
follows,
\begin{equation}
\partial_\mu \phi_\nu -\partial_\nu \phi_\mu = \partial_\mu \tilde{\phi}_\nu
-\partial_\nu \tilde{\phi}_\mu - g s_{\mu \nu}(\tilde{\Sigma}),
\end{equation}
where the second term is concentrated on $\tilde{\Sigma}$ and compensates the
$\delta$-dis\-tri\-bu\-tion on $\tilde{\Sigma}$ that originated when taking the
derivatives of the discontinuous vector field $\tilde{\phi}_\mu(x)$.
 
Because of the multivalued character of the fields, the factors containing the
defects in eqs. (\ref{WYMb}), (\ref{W3d}) become,
\begin{equation}
e^{i\int d^3x\, (\lambda_{\mu} + \frac{g}{2} s_{\mu}) d^{(n)}_{\mu}}=
e^{i\int d^3x\, (\partial_{\mu}\tilde{\phi} +B_\mu )d^{(n)}_{\mu}},
\end{equation}
\begin{equation}
e^{i\int d^4x\, \frac{1}{2}(\lambda_{\mu \nu}+g s_{\mu \nu})d^{(n)}_{\mu \nu}}=
e^{i\int d^4x\, \frac{1}{2}(\partial_\mu \tilde{\phi}_\nu -\partial_\nu
\tilde{\phi}_\mu +B_{\mu \nu})d^{(n)}_{\mu \nu}}
\end{equation}
where we used,
\begin{equation}
\frac{g}{2} \int d^3x\, (s_{\mu}-s_\mu(\tilde{\Sigma}))\, d^{(n)}_{\mu}=2n\pi , 
\end{equation}
\begin{equation}
\frac{g}{2} \int d^4x\, (s_{\mu \nu}-s_{\mu \nu}(\tilde{\Sigma}))\, d^{(n)}_{\mu
\nu}=2n\pi .
\end{equation}
In addition, the implicit constraints in eqs. (\ref{j-map}), (\ref{const-3d})
become,
\begin{equation}
J^c_\mu=\epsilon_{\mu \nu \rho} \partial_\nu (B_{\rho}+\frac{g}{2}\,
[s_{\rho}-s_{\rho}(\tilde{\Sigma})])=\epsilon_{\mu \nu \rho} \partial_\nu
B_{\rho},
\end{equation}
\begin{equation}
J_c^\mu=\frac{1}{2}\epsilon_{\mu \nu \rho \sigma}\partial_\nu (B_{\rho \sigma}+g
[s_{\rho \sigma}-s_{\rho \sigma}(\tilde{\Sigma})])=\frac{1}{2}\epsilon_{\mu \nu
\rho \sigma}\partial_\nu B_{\rho \sigma},
\end{equation}
where we used that the sources $s_\mu$, $s_{\mu \nu}$ are concentrated on
$S({\cal C})$, sharing the same border ${\cal C}$ with $\tilde{\Sigma}$.

Therefore, using the above results when considering multivalued dual fields
$\phi$, $\phi_\mu$, we can represent the Wilson loop in eqs. (\ref{WYMb}),
(\ref{W3d}) according to,
\begin{eqnarray}
\bar{W}({\cal C}) &=& \int [{\cal D}\tilde{\Sigma}][{\cal D}{\cal
F}({\tilde{\Sigma}})]\, e^{-S_c-\int d^3x\, \frac{1}{2}(\partial_\mu
\tilde{\phi} - \frac{g}{2} s_\mu(\tilde{\Sigma})+B_\mu)^2}\times\nonumber \\
&&\times e^{i\int d^3x\, \{ (\epsilon_{\mu \nu \rho} \partial_\nu B_{\rho}-
J^c_\mu ) A^{(n)}_\mu +\lambda_\mu k_{\mu}+(\partial_\mu \tilde{\phi}
+B_\mu)d^{(n)}_{\mu}\}},
\end{eqnarray}
\begin{eqnarray}
\bar{W}({\cal C})
&= &\int [{\cal D}\tilde{\Sigma}][{\cal D}{\cal F}({\tilde{\Sigma}})] \,
e^{-S_c-\int d^4x\, \frac{1}{4}(\partial_\mu \tilde{\phi}_\nu -\partial_\nu
\tilde{\phi}_\mu - g s_{\mu \nu}(\tilde{\Sigma})+B_{\mu \nu})^2}\times\nonumber
\\
&&\times e^{i\int d^4x\, \{ (\frac{1}{2}\epsilon_{\mu \nu \rho \sigma}
\partial_\nu B_{\rho \sigma}-
J^c_\mu ) A^{(n)}_\mu +\frac{1}{2}\lambda_{\mu \nu} k_{\mu
\nu}+\frac{1}{2}(\partial_\mu \tilde{\phi}_\nu -\partial_\nu \tilde{\phi}_\mu
+B_{\mu \nu})d^{(n)}_{\mu \nu}\} },\nonumber \\
\end{eqnarray}
$[{\cal D}{\cal F}({\tilde{\Sigma}})]=[{\cal D}B][{\cal D}\tilde{\phi}]
[{\cal D}\Psi]F^{B}_{gf}F^{\tilde{\phi}}_{gf}$, where $F^{B}_{gf}$ is the part
of the measure fixing the condition for $B_{\mu}$, $B_{\mu \nu}$, and in four
dimensions $F^{\tilde{\phi}}_{gf}$ is the part fixing the condition for
$\tilde{\phi}_\mu$.

In this manner, $\bar{W}({\cal C})$ no longer refers to the particular surface
$S({\cal C})$, initially introduced in the PD representation. In turn, the
path-integral over multivalued fields is equivalent to the integral over all the
surfaces $\tilde{\Sigma}$ with border ${\cal C}$, together with the
path-integral over the  fields $\tilde{\phi}$, $\tilde{\phi}_\mu$, with a given
jump at $\tilde{\Sigma}$.

\section{Conclusions} 
\label{conc}

In this work we have presented a natural framework for discussing possible ideas
underlying confinement and ensembles of defects in $3D$ and $4D$ $SU(2)$
Yang-Mills theory in the continuum.

Initially, we have considered a representation for the Wilson loop average
$\bar{W}$, based on the Petrov-Diakonov representation of the nonabelian Wilson
loop $W$, combined with the Cho-Faddeev-Niemi decomposition of $SU(2)$ gauge
fields, which permits to write the average $\bar{W}$ as a path-integral over
$SU(2)$ mappings. These mappings induce local frames $\hat{n}_a$ in color space,
whose defects represent not only the monopole sector, but also a $Z(2)$ center
vortex sector.

The interesting point is that the integrand of $\bar{W}$ contains an arbitrary
surface $S({\cal C})$, whose 
border is the Wilson loop, originated from the Wess-Zumino term in the
Petrov-Diakonov representation. 
On the other hand, the usual representation for $W({\cal C})$ only refers to
${\cal C}$. Then, the problem is how the representation for $\bar{W}$ can be
worked out so as to implement the independence on the initial choice for
$S({\cal C})$.

In other words, when defects  proliferate, the natural question that arises is
how and under what conditions the surface $S({\cal C})$ can be decoupled, in
favor of its border, or it becomes a Wilson surface variable.

On the other hand, the discussion about how a surface can become observable is a key point to
understanding the possible mechanisms underlying confinement and its associated
properties. 

In ref. \cite{3}, this has been analyzed by means of the peculiar declustering
properties of correlators involving the Wilson loop operator $\hat{W}({\cal C})$
and the disorder operator $\hat{V}(C')$.

In ref. \cite{polyakov1}, the Wilson loop average $\bar{W}$ has been considered
in the context of compact $QED(3)$, and compact $QED(4)$ regularized on the
lattice. There, considering in $3D$ the dual field $\phi$ defined on the
interval $[-\infty, +\infty]$, a representation based on axion fields, with
multivalued action, has been obtained, and a series of approximations led to an
explicit dependence of the resulting $\bar{W}$ on the arbitrary $S({\cal C})$
appearing in its definition. Then, it has been conjectured that this problem
would be resolved if all the branches of the multivalued action were considered
in the calculation, and that this would be equivalent to considering an
integration over all Wilson surfaces,
and dual fields $\tilde{\phi}$ with a given jump at the corresponding surface. 

In this article we have discussed this kind of problem in terms of the
regularity properties imposed on the dual fields by the different ensembles of
defects, and the associated closure properties under large dual transformations.

Our representation for $\bar{W}$ contains an integral over the ensemble of
defects, a path-integral over the diagonal and off-diagonal gluon fields,
including a gauge fixing, and one over dual fields $\lambda_\mu=\partial_\mu
\phi+ B_\mu$, $\lambda_{\mu \nu}=\partial_\mu \phi_\nu -\partial_\nu \phi_\mu +
B_{\mu \nu}$, minimally coupled to the center vortex worldlines or worldsheets,
in three and four dimensions, respectively. 

In terms of the effective action $S_{v,m}$ originated from the ensemble
integration, the effective model for
the Yang-Mills partition function is of the form,
\[ S_{{\rm eff}}=S_{v,m}[\lambda_{\mu}]+ S [\tilde{F}_\mu] + \gamma
\lambda_{\mu} \lambda_{\mu} 
\makebox[.5in]{,}
\tilde{F}_\mu=\epsilon_{\mu \nu \rho}\partial_\nu \lambda_{\rho},
\]
\[
S_{{\rm eff}}=S_{v,m}[\lambda_{\mu \nu}]+  S [\tilde{H}_\mu] + \gamma
\lambda_{\mu \nu} \lambda_{\mu \nu}
\makebox[.5in]{,}
\tilde{H}_\mu=\epsilon_{\mu \nu \rho \sigma}\partial_\nu \lambda_{\rho \sigma},
\]
in $3D$ and $4D$, respectively. 

For example, in $3D$, we have argued that for chains of monopoles attached in
pairs to center vortices, $S_{v,m}$ is naturally associated with a vortex field
$V(x)$, minimally coupled with $\lambda_\mu$,
displaying a local $Z(2)$ symmetry. This symmetry is also present in the second
term $S[\tilde{F}_\mu]$. However, because of the last term $\lambda_\mu
\lambda_\mu$, the $Z(2)$ symmetry in $S_{{\rm eff}}$ is only global, and the
effective model is expected to be a generalization of the well-known vortex
model of ref. \cite{3}.

Moreover, in a phase where the global $Z(2)$ is spontaneously broken, the
effective theory contains domain walls, a topological structure whose existence
depends on the consideration of well-behaved continuous fields $V(x)$,
$\lambda_\mu$. 
Then, the change of variables associated with the local $Z(2)$ transformations
$\phi \rightarrow \phi + \omega^{(3)}$, adding to $\lambda_\mu$ a source
localized on a closed Wilson surface, cannot even be accepted in this case, as
these transformations are not closed. On the contrary, if there is no SSB, the
regularity requirement is no longer valid, and this change of variables becomes
acceptable. 

Similarly for monopole chains in $4D$, the $\lambda_{\mu \nu} \lambda_{\mu \nu}$
term in $S_{{\rm eff}}$  would be the noninvariant part under large dual
transformations $\phi_\mu \rightarrow \phi_\mu +\partial_\mu \omega^{(4)}$,
adding a source localized on a closed Wilson surface $\partial \vartheta$. Here
the discussion about a possible topological structure for the effective theory
is highly nontrivial, as the vortex field $V(x)$ in $3D$ is replaced by a loop
variable $V(C)$. Nevertheless, we can asssume that different phases could exist,
where the associated regularity requirements on $\lambda_{\mu \nu}$ could lead
to consider the changes of variables $\phi_\mu \rightarrow \phi_\mu +
\partial_\mu \omega^{(4)}$ as acceptable or not. 

For example, this discussion already occurs in $4D$ when looking at the monopole
part of $S_{v,m}$, in the simpler situation where monopoles are uncorrelated
with center vortices. As is well-known, this part is typically   represented by
a complex field $\Phi(x)$, minimally coupled with $\phi_\mu$. In a phase where
the dual $U(1)$ is spontaneously broken, the model has a topological structure,
whose existence depends on the consideration of well-behaved continuous fields
$\Phi(x)$, $\lambda_{\mu \nu}$, thus precluding the $\phi_\mu \rightarrow
\phi_\mu +\partial_\mu \omega^{(4)}$ transformations. 

In canonical language, this corresponds to the fact that an order parameter must
be invariant under regular gauge transformations. For the condition
$\partial_\mu \phi_\mu =0$, such an order parameter turns out to be variant not
only under spacetime independent phase transformations, but also under
multivalued $\omega^{(4)}$-transformations. Then, if the global $U(1)$ is
spontaneously broken, the large dual transformations will also display SSB.

For these reasons, in the last part of this work, we were led to analyze the
representation for $\bar{W}$ in two possible scenarios, before considering an
effective model for the ensemble integration.

In the representation of $\bar{W}$ in $3D$, we have discussed two alternatives
for the class of fields $\lambda_\mu$. They are generated by $\phi^{{\rm I}}$,
general single-valued fields defined on the interval $[-\infty, +\infty]$, or by
the fields $\phi^{{\rm II}}$, multivalued when we go around the Wilson loop
${\cal C}$.  While in the former case changes of variables $\phi^{{\rm I}}
\rightarrow \phi^{{\rm I}}+\omega^{(3)}$ are acceptable, in the latter, the
addition of $\phi^{{\rm II}}$ with $\omega^{(3)}$ is meaningless. 

These alternatives have been generalized to $4D$, where the class of fields
$\lambda_{\mu \nu}$ can be generated by two types of fields. The first type is
closed under the transformation $\phi^I_\mu \to \phi^I_\mu +\partial_\mu
\omega^{(4)}$, with $\omega^{(4)}$ a multivalued phase when we go around
$\partial \vartheta$. For the second type, this transformation does not make any
sense, as the fields cannot be globally defined on the closed surfaces linked by
the Wilson loop ${\cal C}$.

In general, if in $3D$ or $4D$ the required fields are type I, we have shown
that it is possible to perform changes of variables in the representation for
$\bar{W}$ so as to decouple the Wilson surface $S({\cal C})$. 

In the second case, the integral over type II multivalued fields were replaced
by an integral over all possible surfaces $\tilde{\Sigma}$ whose border is
${\cal C}$, and dual fields $\tilde{\phi}$, $\tilde{\phi}_\mu$, functions of
point $x$ on the Euclidean spacetime, with an appropriate jump at
$\tilde{\Sigma}$. In this manner, any reference to the initial arbitrary
$S({\cal C})$ also disappeared, but in a different way; the initial surface has
become a Wilson surface variable.

Summarizing, for $SU(2)$ Yang-Mills theories, we introduced a framework to 
discuss the coupling between gauge fields containing defects, surfaces attached
to the Wilson loop, and dual fields. We have discussed some effective models, the implied
regularity requirements and the associated inequivalent manners to represent the Wilson 
loop without reference to the initial Wilson surface considered.
This general framework could prove useful as a starting point to understand the
promising scenario associated with correlated monopoles and center vortices in
continuum $4D$ Yang-Mills theories.

\section{Acknowledgements}

The Conselho Nacional de Desenvolvimento Cient\'{\i}fico e Tecnol\'{o}gico
(CNPq-Brazil) and the Funda{\c {c}}{\~{a}}o de Amparo
{\`{a}} Pesquisa do Estado do Rio de Janeiro (FAPERJ) are acknowledged for the
financial support.


\end{document}